%% file: GBCNetworksArxiv.tex
\documentclass[12pt]{article}
\usepackage{latexsym,graphics,graphicx,amssymb,amsmath, amstext, amsfonts, amsthm, pifont, color,  subfig}
\input{newcommand}
\makeatletter

\author{Sreeram Kannan, Adnan Raja,  and Pramod Viswanath }

\title{Approximately Optimal Wireless Broadcasting}

\begin{document}
\maketitle

\begin{abstract}

We study a wireless broadcast network, where a single source
reliably communicates independent messages to multiple destinations,
with the aid of relays and cooperation between destinations. The wireless nature of the medium is
captured by the {\em broadcast} nature of transmissions as well as
the {\em superposition} of all transmit signals plus independent Gaussian
noise at the received signal at any radio. We propose a scheme that can achieve rate tuples
within a constant gap away from the cut-set bound, where the constant is independent of channel coefficients and power constraints.

The proposed scheme operates in two steps. The {\em inner} code, in which the relays perform a quantize-and-encode operation, is constructed by lifting a
scheme designed for a corresponding discrete superposition network. The {\em outer} code is a Marton code for the non-Gaussian vector broadcast channel induced by the relaying scheme, and is constructed by adopting a ``{\em receiver-centric}'' viewpoint.

\end{abstract}

\section{Introduction}\label{sec:intro}

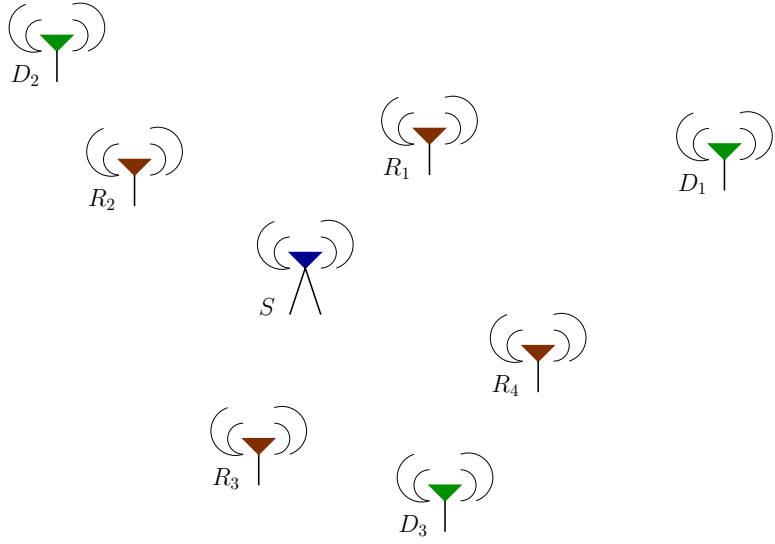
\begin{figure}[htb]
\begin{center}
\scalebox{0.65}{\input{figs/network.pstex_t}}
\end{center}
\caption{A wireless broadcast network} \label{fig-setting}
\end{figure}

The scenario of study in this paper is depicted in
Figure~\ref{fig-setting}. A single source node is reliably
communicating independent messages to multiple destination nodes
using the help of multiple relay nodes. In the example of a cellular
system, the setting represents downlink communication where the
base-station is transmitting to multiple terminals with the
potential help of relay stations. Note that some of the
terminals can themselves act as relays. A special instance of our
setting is the following: only the source node and multiple
destinations (i.e., no relays) are present. Since we have allowed
the ability to transmit and receive at all nodes, this special
instance models the downlink of a cellular system with the
destinations having the ability to cooperate among themselves, which
has been studied in \cite{LV07}.

We consider the canonical Gaussian channel model among the various
nodes in the network: time is discrete and synchronized among all
the nodes. Denoting the baseband transmit symbol (a complex number)
of node $k$ at time $m$ by $x_k[m]$, the average transmit power
constraint at each node implies that, \beq
  \sum_{m=1}^T |x_k[m]|^2 \leq TP_k,
\eeq where $T$ is the time period over which the communication
occurs. At each time $m$, we have the received signal at any node
$\ell$ \beq y_\ell[m] = \sum_{k\neq \ell} h_{k\ell}[m] x_k[m] +
z_\ell[m]. \label{eq:Gausschannel} \eeq Here $\lbr z_\ell[m] \rbr_m$ is i.i.d.\ Gaussian
noise and independent across the different nodes $\ell$. The channel
attenuation $h_{k\ell}$ between a pair of nodes $(k,\ell)$ is
supposed to be constant over the time scale of communication. Note
that by normalizing the channel attenuation $h_{k\ell}$, without
loss of generality, we will assume unit average power constraints at
each node, i.e.~$P_{k}= 1$ and also the variance of $z_\ell[m]$ to
be $1$. We suppose full duplex mode of operation for the most part,
while discussing the implications of half duplex mode later in the
paper. We suppose single antenna at each node and leave the
discussion with multiple antennas for a later part of the paper.  We
will begin with the supposition that these channel attenuations are
known to all the nodes in the network, and revisit this requirement
later.

Let $\mc{V}$ denote the set of all nodes in the network. A
$(2^{TR_1},\ldots,2^{TR_J},T)$ coding scheme for the broadcast
network, with source node $S$ and destination nodes
$D_1,\ldots,D_J$, which communicates over $T$ time instants  is
comprised of the following \ben
\item Independent random variables $W_{i}$ which are distributed uniformly on $[2^{TR_i}]$ for $i=1,\ldots,J$ respectively. $W_{i}$ denotes the message intended for destination $D_{i}$.
\item The source mapping, \beqa f_{S}: (W_{1} \times \ldots \times W_{J}) \rightarrow \mc{X}_S^T.  \eeqa
%\item The source mapping for $t \in [T]$, \beqa f_{S,t}: (W_{1} \times \ldots \times W_{J}, \mc{Y}_{S}^{t-1}) \rightarrow \mc{X}_S^t.  \eeqa
\item The relay mappings for each $v \in \mc{V} \backslash \lbr S \rbr$ and $t \in [T]$, \beqa f_{v,t} : \mc{Y}_v^{t-1} \rightarrow \mc{X}_v. \label{eq:genRelMap} \eeqa
\item The decoding map at destination $D_i$, \beqa g_{D_i} : \mc{Y}_{D_i}^T \rightarrow \hat{W}_{i}. \eeqa
\een The probability of error for destination $i$ under this coding
scheme is given by \beqa P_e^i  & \df & \Pr \{ \hat{W}_{i} \neq W_i
\}. \eeqa A rate tuple $(R_1,R_2,...,R_J)$, where $R_{i}$ is the
rate of communication in bits per unit time for destination $D_{i}$,
is said to be achievable if for any $\epsilon >0 $, there exists a
$(2^{TR_1},2^{TR_2},...,2^{TR_J},T)$ scheme that achieves a
probability of error lesser than $\epsilon$ for all nodes, i.e.,
$\max_i P_e^i \leq \epsilon$. The capacity region $\mc{C}$ is the set
of all achievable rates.

The following is the well known cut-set upper bound to the rate
tuples of reliable communication \cite{E81,elemIT}: denoting the set
of all nodes by $\mc{V}$; and for all subsets $\mc{J}\subseteq [J]$, where $[J]$ denotes the set $\lbr 1,\ldots,J \rbr$, denoting $\Lambda_{\mc{J}}$ to be the collection
of all subsets $\Omega \subset \mc{V}$ such that the source nodes
$S\in\Omega$ and a subset $\mc{J}\subseteq [1:J]$ of destinations
$\mc{D}_{\mc{J}} \in \Omega^c$; we have that if $(R_1,\ldots ,R_J)$
is achievable, $\forall\mc{J}$, there is a joint
distribution $p\lp \lbr X_{v} | v \in \mc{V}\rbr \rp$ (denoted by Q)
such that \beq  R_{\mc{J}} \leq \bar{C}_{\mc{J}}(Q) \df \min_{\Omega
\in \Lambda_{\mc{J}}} I \lp Y_{\Omega^{c}}; X_{{\Omega}} |
X_{\Omega^{c}} \rp, \eeq where $R_{\mc{J}} \df \sum_{j\in {\mc{J}}}
R_j$.

Let $ \bar{\mc{C}}(Q)$ denote the set of all rate tuples that
satisfy the cut-set upper bound for a given joint distribution $Q$, and $\bar{\mc{C}}$ denote the cut-set bound:
\beqa  \bar{\mc{C}}(Q) & \df & \{ (R_1,...,R_J): R_{\mc{J}} \leq \bar{C}_{\mc{J}}(Q) \ \forall \mc{J} \subseteq[1:J] \}  \\
 \bar{\mc{C}} & \df & \text{conv} \lp \bigcup_{\{Q: \mathbb{E} |X_v|^2 \leq 1 \}}  \bar{\mc{C}}(Q) \rp, \eeqa
where $\text{conv}\lp\cdot\rp$ denotes the convex hull of the region.

Our main result is the following. \bthm \label{thm:main} For the
wireless broadcast network, a rate vector $(R_1,\ldots ,R_J)$ is
achievable if $\forall~\mc{J}$, \beq (R_1+k,\ldots,R_J+k)
\in \bar{\mc{C}} \eeq
for some constant $k$, which depends only on the number of nodes,
and not on the channel coefficients, and  $k = O(|\mc{V}| \log |\mc{V}| )$. \ethm

With a single destination node ($J = 1$), the scenario reduces to
the classical Gaussian (unicast) relay channel. Pioneering work of
\cite{ADT09} has obtained an approximate characterization of the
capacity for this scenario.
In recent work,  \cite{AK10,LKEC10} derive similar approximation results with different
coding schemes. In particular, \cite{ADT09,AK10} take a two-step
approach in their coding scheme: first, they develop deterministic
models that approximate the Gaussian channel and next, they
construct the codes for the Gaussian channel based on the codes for
the corresponding deterministic channel approximation. The specific
approaches adopted in the actual deterministic approximation and
moving from the codes for the deterministic channel to the Gaussian
one are different between \cite{ADT09} and \cite{AK10}; as such, we
follow the approach of \cite{AK10} in our proof of the main result.

The proposed scheme operates in two steps. The {\em inner} code, in which the relays essentially perform a quantize-and-encode operation, is constructed by lifting a
scheme designed for a corresponding discrete superposition network.
This induces a vector broadcast channel between

The {\em outer} code is essentially a Marton code (\cite{M79,EM81}) for the broadcast channel induced by the relaying scheme, and is constructed based on a ``receiver-centric'' viewpoint.

The rest of the paper is organized as follows. In Section
\ref{sec:deterministic}, we give a coding scheme and establish an
achievable rate region for deterministic broadcast networks. In
Section \ref{sec:Gaussian}, we prove Theorem \ref{thm:main} by
giving a coding scheme for the wireless broadcast network. In order
to do so, we use the discrete superposition network, which is an
approximation to the wireless network, and the ``lift'' the scheme from the discrete superposition network to the Gaussian network. In Section~\ref{sec:discussion}, we discuss various aspects of the proposed scheme, primarily the reciprocity in the context of linear deterministic and Gaussian networks, and the channel state information required. In Section~\ref{sec:generalization}, various generalizations of the scheme are provided, for half-duplex networks, for networks with multiple antenna and for broadcast wireless networks, where some set of nodes demand the same information and other nodes demand independent information.

\section{Deterministic Broadcast Networks} \label{sec:deterministic}

In the deterministic network model, the received signal at each node
is a deterministic function of the received signals. \beq
y_{\ell}[m] = g_{\ell}\lp\lbr x_k[m] \rbr_{k\neq\ell}\rp.
\label{eq:Detchannelmodel} \eeq The input and output alphabet sets,
$\mc{X}_{k}$'s and $\mc{Y}_{\ell}$'s respectively, are assumed to be
finite sets.

As before, we have the following cut-set upper bound to the rate
tuples of reliable communication \cite{E81,elemIT}: if $(R_1,\ldots
,R_J)$ is achievable, then $\forall\mc{J}\subseteq[1:J]$, there is a
joint distribution $p\lp \lbr X_{v} | v \in \mc{V}\rbr \rp$ (denoted
by Q) such that \beq  R_{\mc{J}} \leq \bar{C}_{\mc{J}}(Q) \df
\min_{\Omega \in \Lambda_{\mc{J}}} I \lp Y_{\Omega^{c}};
X_{{\Omega}} | X_{\Omega^{c}} \rp. \eeq

We prove the following achievability result for the deterministic
channel.
\begin{theorem} \label{thm:capGenDet}
For the deterministic broadcast network, a rate vector $(R_1,\ldots
,R_J)$ is achievable if $\forall~\mc{J}$, there is a product
distribution $\prod\limits_{v\in\mc{V}} p(X_{v})$ (denoted by
$Q_{p}$) such that \beq R_{\mc{J}} \leq \bar{C}_{\mc{J}}(Q_{p}).
\eeq
\end{theorem}

\begin{figure}[tb]
\begin{center}
\scalebox{0.65}{\input{figs/BC-Relay.pstex_t}}
\end{center}
\caption{A Layered broadcast-relay network} \label{fig-BCN}
\end{figure}
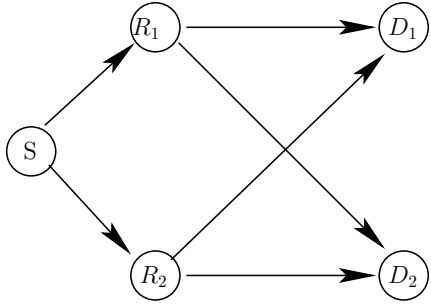

\begin{remark} Aref networks, i.e., deterministic broadcast networks where each node can receive information from every incoming edge separately, were studied in \cite{VasKor}. It is shown there that cut-set bound can be achieved for these networks. This result can also be recovered from Theorem~\ref{thm:capGenDet} by observing that product form distributions optimize the cut-set bound. It should be noted, however, that the scheme proposed in \cite{VasKor} is a separation-based scheme whereas the scheme proposed here is not. \end{remark}

\bpf We prove Theorem \ref{thm:capGenDet} for the layered network
here. The arguments can then be extended to the general network by
using time expansion as done in \cite{ADT09}. A network is called a
{\em L-layered network} if the set of vertices $\mc{V}$ can be
partitioned into $L$ disjoint sets, such that only the source node
$S$ is in the first layer and the $J$ destination nodes are in the
$L$-th layer. The nodes in the intermediate layers are relaying
nodes. The received signal at the nodes in the $l+1$-th layer only
depend on the transmitted signals at the nodes in the $l$-th layer.
This dependency is often represented by edges connecting the nodes
from the $l$-th layer to the $(l+1)$-th layer. An example of a
layered broadcast network is shown in Fig.~\ref{fig-BCN}. The
advantage of working with a layered network is that we can view the
information as propagating from one layer to the next without
getting intertwined.

\subsection{Outline of Coding Scheme}

\begin{figure}[htb]
\begin{center}
\scalebox{0.65}{\input{figs/codingBlocks.pstex_t}}
\end{center}
\caption{Coding scheme} \label{fig-CB}
\end{figure}
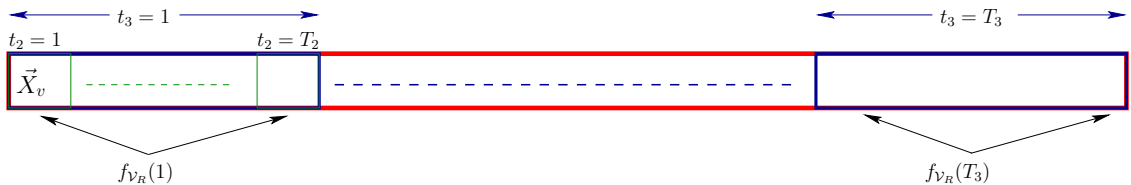

The coding scheme operates over three levels of nested blocks as
shown in Figure \ref{fig-CB}. If $T$ is the total time period of
communication, then $T=T_{1}T_{2}T_{3}$. The innermost level is
level-1 and consists of $T_{1}$ time instants. $T_{2}$ such
level-$1$ blocks constitutes a level-$2$ block. The overall coding
scheme operates over $T_{3}$ level-$2$ blocks.  Our coding scheme
comprises of the following: \beit
\item The  {\em message $W_{i}$} is broken into $T_{3}$ independent sub-messages $W_{i}(1),\ldots,W_{i}(T_{3})$, where each sub-message is encoded over a particular level-$2$ block.
\item The {\em relay mappings} are done at the level-$1$ block. Every node blocks up $T_{1}$ received symbols and maps it to $T_{1}$ transmit symbols. This relay mapping is fixed for the duration of a level-$2$ block and it essentially creates an end-to-end deterministic broadcast channel at the level-$2$ block.
\item  The {\em encoding at the source} is done on the induced end-to-end deterministic broadcast channel for each level-$2$ block. Across different level-$2$ blocks, the mappings are generated in an i.i.d.~manner and corresponding broadcast schemes are used for the different end-to-end broadcast channel induced by this operation. This is very much like a fading broadcast channel. Here the random fading is introduced by the relay mappings. This is merely a proof technique which allows us to average the performance over random relay mappings.
\item  The {\em destinations} decode the sub-message corresponding to each level-$2$ independently and sequentially.
\eeit

\subsection{Coding Scheme in Detail} \label{sec:codingSchemeDN}

Throughout the discussion below, we fix a particular product distribution $Q_{p}$, which is then used to describe a random ensemble of coding operations.

\subsubsection{Relay mappings} \label{sec:relayMap}

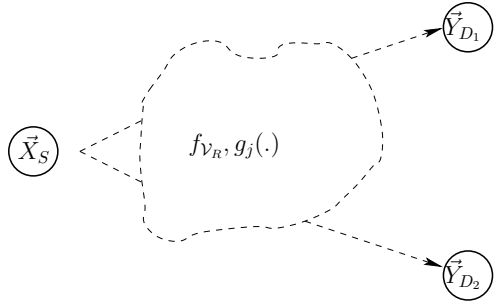
\begin{figure}[htb]
\begin{center}
\scalebox{0.65}{\input{figs/outerCode-BC.pstex_t}}
\end{center}
\caption{Effective end-2-end deterministic broadcast channel created
by a level-$1$ code.} \label{fig-OC}
\end{figure}

In the proposed scheme, the relays operate over
level-$1$ blocks, i.e., each relay transmits in a level-$1$ block using only the information from the previous
received level-$1$ block. We will use $\vec{x}_{r} \df x_{r}^{T_{1}}$ and
$\vec{y}_{r} \df y_{r}^{T_{1}}$ to denote the transmit and receive
block at any relay node $r \in \mc{V}_{R}$, where $\mc{V}_{R}$ is
the set of all relay nodes. The mapping at the relay node denoted by
 \beq f_{r} (t_3) :  \mc{Y}_{r}^{T_1} \rightarrow
\mc{X}_{r}^{T_1}, \eeq
is random and generated  i.i.d.~ from the distribution $p(X_{r})$, i.e., $\forall y_r^{T_1} \in \mc{Y}_r^{T_1}$, generate $x_r^{T_1}$ i.i.d. from $p(X_{r})$.
Note that the relay mapping is only a function of $t_{3}$ i.e., it
is fixed across all level-$1$ blocks in a given level-$2$ block.
Across different level-$2$ blocks, i.e., for each $t_3$, it is generated in an
i.i.d.~manner. As mentioned earlier, this induces an end-to-end
broadcast channel as shown in Figure \ref{fig-OC}; a different one
across every level-$2$ block.

\subsubsection{Source Mappings \label{sec:sourceMap}}

The capacity of the deterministic broadcast channel is well known
(\cite{M79,Pinsker}). For our coding scheme over the
induced deterministic broadcast channel of Figure \ref{fig-OC}, we
use the coding scheme similar to the one described for the
deterministic broadcast channel in \cite{EM81}, which we refer to henceforth as the
``Marton code". The {\em source codebook} for
the $t_{3}$-th level-$2$ block, which maps the message $W_i(t_{3})
\in [2^{T_1T_2 R_i(t_3)}], i=1,2,...,J$ to transmit symbol block
$\vec{x}_S^{T_2}$, is described below.

Given the random vector $\vec{X}_{S}$ which is distributed as
$p({\vec{X}_S}) = \prod p(X_{S})$, the channel and the relay mapping
induce the joint distribution over the random variables $\lp
\vec{X}_S,\vec{Y}_{\mc{V}}\rp$. Create auxiliary random variables $\vec{U}_{D_i}$ such that $p_{\vec{X},\vec{U}_{D_1},\vec{U}_{D_2},...,\vec{U}_{D_J}}$ is the same as $p_{\vec{X},\vec{Y}_{D_1},\vec{Y}_{D_2},...,\vec{Y}_{D_J}}$.

The set $\mc{T}_{\delta}^{T_2}(\vec{U}_{D_{i}})$ of all typical
$\vec{u}_{D_{i}}^{T_2}$ are binned into $2^{T_1 T_2 R_i(t_3)}$ bins,
where each bin index corresponds to a message, for $i=1,2,...,J$.
For each vector
$(\vec{u}_{D_{1}}^{T_2},\ldots,\vec{u}_{D_{J}}^{T_2}) \in
\mc{T}_{\delta}^{T_2}(\vec{Y}_{D_{1}},\ldots,\vec{Y}_{D_{J}})$,
there exists a sequence $\vec{x}_S^{T_2}
(\vec{u}_{D_{1}}^{T_2},\ldots,\vec{u}_{D_{J}}^{T_2})$, since the channel is deterministic, such that
$(\vec{x}_S^{T_2},\vec{u}_{D_{1}}^{T_2},\ldots,\vec{u}_{D_{J}}^{T_2}) \in
\mc{T}_{\delta}^{T_2}(\vec{X}_S,\vec{Y}_{D_{1}},\ldots,\vec{Y}_{D_{J}})$.
This specifies the codebook for the given level-$2$ block $t_{3}$.
Similar codebooks are generated, statistically independently, for all the $T_{3}$ level-$2$
blocks.

\subsubsection{Encoding}

The messages $W_i \in [2^{T_1 T_2 T_3 R_i}]$ are split into
sub-messages $W_i(t_3) \in [2^{T_1 T_2 R_i(t_3)}]$ such that \beq
\sum_{t_3}R_i(t_3) = R_i T_3. \label{eq:rateSplit} \eeq For the
$t_{3}$-th level-$2$ block, the messages to be transmitted are given
by $\lp W_{1}(t_3),\ldots,W_{J}(t_3) \rp$ for the $J$ destinations
respectively. To transmit the message, the source looks at the
codebook for the level-$2$ block $t_{3}$ and tries to find a vector
$\lp \vec{u}_1^{T_{2}},\ldots,\vec{u}_J^{T_{2}} \rp \in
\mc{T}_{\delta}^{T_2}({\vec{U}_1,\ldots,\vec{U}_J} )$ such that
$\vec{u}_i^{T_{2}}$ is also in the bin with index $W_{i}(t_{3})$. If
the source can find such a vector, it transmits
$\vec{x}_s^{T_2}(\vec{u}_1^{T_{2}},\ldots,\vec{u}_J^{T_{2}})$. If
the source cannot find such a sequence it transmits a random
sequence.

\subsubsection{Decoding}
At the end of $t_3$-th level-$2$ block, the destination $D_i$
decodes the transmitted message $W_i(t_3)$. The destination tries to
find the bin in which the received level-$2$ block
$\vec{y}^{T_2}_{D_i} (t_3)$ falls and decodes that bin index as the
transmitted message.

\subsection{Performance Analysis} \label{sec:perfAnalDN}

We begin with identifying rate constraints for the $t_{3}$-th level-$2$
block, so that arbitrarily small probability of error can be
achieved for decoding the message at this block. From the coding
theorem for the deterministic broadcast channel (\cite{M79}, Theorem
3), we know that as long as $R_{i}(t_{3})$ satisfies, for $i=1,\ldots,J$
and $\forall \mc{J} \subseteq \lbr 1,\ldots,J \rbr$, \beq
R_{\mc{J}}(t_3) \leq \frac{1}{T_1}
H(\vec{Y}_{D_\mc{J}}|F_{\mc{V}_{\mc{R}}}=f_{\mc{V}_{\mc{R}}}(t_3)),
\label{eq:level2_perf} \eeq the probability of error can be made
arbitrarily small by choosing a large enough $T_{2}$.
The overall rate $R_{i}$ is given by \eqref{eq:rateSplit}. Therefore
the rate tuple $\lp R_1,\ldots,R_J \rp$ satisfies \beq R_{\mc{J}}
\leq \frac{1}{T_{1}T_3}  \sum_{t_3=1}^{T_3}
H(\vec{Y}_{D_\mc{J}}|F_{\mc{V}_{\mc{R}}}=f_{\mc{V}_{\mc{R}}}(t_3)),
\forall \  \mc{J} \subseteq \lbr 1,\ldots,J \rbr. \label{eq:condRi}
\eeq By the strong law of large numbers, as $T_3 \rightarrow \infty$,
we have
\beqa \frac{1}{T_1 T_3}  \sum_{t_3=1}^{T_3} H(\vec{Y}_{D_\mc{J}}|F_{\mc{V}_{\mc{R}}}=f_{\mc{V}_{\mc{R}}}(t_3)) & \stackrel{\text{a.s.}}{\rightarrow} & \frac{1}{T_1}\mb{E} H(\vec{Y}_{D_\mc{J}}|F_{\mc{V}_{\mc{R}}}=f_{\mc{V}_{\mc{R}}}(t_3)) \nn \\
& = & \frac{1}{T_1} H(\vec{Y}_{D_\mc{J}}|F_{\mc{V}_{\mc{R}}}).
\label{eq:SLLN}\eeqa Next, we relate the expression in
\eqref{eq:SLLN} to the cut-sets using the following lemma: \blem
\label{lem:mI} Given arbitrary $\epsilon > 0,~\exists~T_{1}~$ s.t.,
\begin{align}
H(\vec{Y}_{D_{\mc{J}}}|F_{\mc{V}_{R}}) = H(Y_{D_{\mc{J}}}^{T_{1}}|F_{\mc{V}_{R}}) \geq T_{1}&(\bar{C}_{\mc{J}}(Q_{p}) - \epsilon),  \label{eq:inner_code} \\
& \forall \  \mc{J} \subseteq \lbr 1,\ldots,J \rbr. \nn
\end{align}
\elem \bpf See Appendix \ref{app:A}. \epf Using \eqref{eq:condRi},
\eqref{eq:SLLN} and Lemma \ref{lem:mI}, we conclude that for any rate
tuple satisfying \beq R_{\mc{J}}  \leq  \bar{C}_{\mc{J}}(Q_{p}) -
\epsilon, \eeq arbitrarily small probability of error is achieved.
The proof is finally completed by allowing time-sharing between the
coding schemes for all product distributions $Q_{p}$. \epf

\section{Gaussian Broadcast Networks (Proof of Theorem \ref{thm:main}) } \label{sec:Gaussian}

\subsection{Layered Network}

As for the deterministic network, we will first consider a layered
network. For the deterministic network, we first used the random
relay mappings to create an effective vector broadcast channel. Then we
used the Marton scheme with a specific choice of auxiliary random variables, which is optimal for the deterministic
broadcast channel. For the Gaussian network, while it is possible to do
the inner code similarly and induce a broadcast channel, this is a vector non-Gaussian broadcast channel for which it is unknown whether a Marton scheme can achieve any rate within a constant gap of the cut-set  bound.
To deal with this issue, we convert the Gaussian
network into a deterministic network, for which we can design the
germane code and then appropriately ``lift" the code from the deterministic to
the Gaussian network.

The procedure is outlined as follows:

\ben
\item Given the Gaussian broadcast network, we construct a corresponding deterministic superposition network (DSN).
The cut-set bound of the DSN approximates the cut-set bound of the corresponding Gaussian network to within a factor $N \log N$. Further,
 the DSN is deterministic and thus the scheme in Theorem~\ref{thm:capGenDet} can be used for the DSN to achieve the
 cut-set bound evaluated under product-form distributions. The details are provided in Sec~\ref{sec:DSN}.

\item We then prune a natural coding scheme $\mc{P}(\kappa)$ for the DSN, such that the rate is reduced by a
factor $N \kappa$. The details are in Sec.~\ref{sec:pruned_scheme}.

\item Finally, we show that for an appropriate choice of $\kappa = O(\log N)$, the pruned coding scheme on the DSN can be {\em emulated}
 on the Gaussian network, because each node in the Gaussian network can decode the corresponding received vector in the DSN. The details are given in Sec.~\ref{sec:lifting}. Therefore, the rate achieved in the Gaussian network by the proposed scheme achieves within a gap of $O ( N \log N)$ of the cut-set bound.

\een

\subsubsection{ Discrete Superposition Network (DSN) } \label{sec:DSN}
Given a Gaussian network, we construct a discrete superposition network.
This network is essentially a truncated noiseless version of the Gaussian model. Further, the input
in this model is restricted to a finite set.

Corresponding to the channel model for the Gaussian network given by \eqref{eq:Gausschannel}, the received signal in the DSN is given by
\beq  y_\ell[m] = \Lbr \sum_{k\neq \ell} h_{k\ell}[m] x_k[m] \Rbr,  \eeq
where $\Lbr\cdot\Rbr$ lies in $\mathbb{Z} + \imath \mathbb{Z}$ and
corresponds to rounding the real and imaginary parts of the complex number to the nearest integer.
The transmit symbols $x_{k}[m]$ are restricted to a discrete and finite complex valued set.
This defines the DSN. Note that our model is very similar to the truncated model in \cite{ADT09} and to the model described in \cite{AK10}.

Next, the following lemma relates the cut-set upper bound of the Gaussian
network, $\bar{\mc{C}}^{\text{Gauss}}$, to the cut-set of
the DSN under product distribution.

    \blem \label{lem:product_form}   There exists a $Q_{p}$ for the DSN such that
    \beq  \bar{\mc{C}}^{\text{Gauss}}  \subseteq  \bar{\mc{C}}^{\text{DSN}}(Q_{p}) +k_{2} (1,1,...,1), \eeq
    where $k_2 = O(|\mc{V}|)$.
    \elem

\bpf See Appendix \ref{app:DSNapprox}. \epf

Note that since the DSN is a deterministic network, we have the
following corollary of Theorem~\ref{thm:capGenDet}. \bcor For the
DSN,  a rate vector $(R_1,\ldots ,R_J)$ is achievable if there is a
product distribution $Q_{p}$  such that, $\forall~\mc{J}$, \beq
R_{\mc{J}} \leq \bar{C}_{\mc{J}}^{\text{DSN}}(Q_{p}). \eeq \ecor
We observe that, here $Q_{p}$ is product distribution over the {\em finite} input
alphabet set of the DSN.

\subsubsection{Pruned Coding Scheme $\mc{P}(\kappa)$ for the DSN} \label{sec:pruned_scheme}

The pruned coding scheme described next is along the lines of the
ideas developed in \cite{AK10}. The main idea there is that a coding
scheme from the DSN can be used in the Gaussian network, if at every
node, the received vector at any node in the DSN can be decoded from the
corresponding received vector in the Gaussian network. This will not,
in general, be true for any scheme in the DSN. Therefore, to get the source codebook for the Gaussian network, the
codewords transmitted by the source in the DSN are pruned to a
smaller set in such a way that the received
vector at any node in the DSN can be decoded from the received vector in the Gaussian network. In this section,
we will demonstrate how to construct a pruned coding scheme
$\mc{P}(\kappa)$, for which the rate is reduced by a constant $N
\kappa$. In Sec.~\ref{sec:lifting}, we will show how to choose the
parameter $\kappa$ such that the Gaussian network can {\em emulate} the
DSN.

Our coding scheme for the deterministic network described in Section
\ref{sec:codingSchemeDN} involved three levels. Level-$1$ involved
relay mappings over a level-$1$ block of $T_1$ time symbols defined
by $f_{\mc{V}_R}$. Then the level-$2$ code involved a Marton scheme
for this fixed $f_{\mc{V}_R}$. This code was over a level-$2$ block,
which comprised of $T_2$ level-$1$ blocks. Then at level-$3$, we
repeated this Marton code over $T_3$ level-$2$ blocks, where for
each super-block the relay mapping $f_{\mc{V}_R}$ was generated
i.i.d.~using $F_{\mc{V}}$. For the Gaussian network, we will
maintain the {\em same level-$1$ and level-$3$ codes}, but will
modify the Marton code by pruning the set of codewords transmitted
by the source. This is described in detail next.

{\em Pruning the Level-$2$ (Marton) code:} We now specify the
operation during the $t_3$-th level-$2$ block. In this level-$2$
block we need to send the messages $W_i \in [2^{T_1T_2 R_i(t_3)}]$
at rates $R_i(t_3)$, for $i=1,\ldots,J$. As before, since the the
relays maintain the same mapping $f_{\mc{V}_R}(t_3)$ for the
duration of this level-$2$ block, we have an effective end-to-end
deterministic channel as shown in Figure \ref{fig-OC}.

As before, given the random variable at the source $\vec{X}_S = X_S^{T_1}$ which
is distributed as $p(\vec{X}_S) = \prod p(X_S)$, the fixed relay
mapping and the channel induces the joint distribution on
$(\vec{X}_S,\vec{Y}_{\mc{V}})$. Let
$\mc{T}^{T_2}_{\delta} (\vec{Y}_v)$ denote the set of all typical
received block vectors at node $v$.

Previously, we created the codebook by binning the typical sets
$\mc{T}^{T_2}_{\delta} (\vec{Y}_{D_{i}})$. But now, we prune
this set before binning. This will, in effect, lead to pruning the
set of typical vectors $\mc{T}^{T_2}_{\delta} (\vec{Y}_v)$ at all
nodes.
To do this pruning, we fix a constant $\kappa$, which will be defined in
Section~\ref{sec:lifting}. At every node $v$, we pick a random
$2^{-T_1 T_2 \kappa}$ fraction of $\mc{T}^{T_2}_{\delta}
(\vec{Y}_v)$ and call this subset $\mf{S}_v$.
We then define, \beq \mf{Z}_{v} \df \{\vec{y}^{T_2}_{v}: \forall i, \ \exists
\vec{y}_i^{T_2} \in \mf{S}_i, \text{ s.t.}~\vec{y}_\mc{V}^{T_2} \in
\mc{T}_{\delta}^{T_2}(\vec{Y}_\mc{V}) \}. \eeq We will prune our codebook so that only codewords in $\mf{Z}_{v}$ are typically received.

Following the Marton code for the
deterministic channel, we bin the set of all $\mf{Z}_{D_i}$, the typically received vectors at the destination, into
$2^{T_1T_2R_i(t_3)}$ bins, for $i=1,\ldots,J$. For a given message pair $W_1,...,W_J$, we select a jointly typical $\vec{y}_{D_1},...,\vec{y}_{D_J} \in \mc{T}^{T_2}_{\delta}(\mf{Z}_{D_1},...,\mf{Z}_{D_J})$  with $\vec{y}_{D_i}$ belonging to the bin corresponding to the message $W_i$, and transmit the $\vec{x}^{T_2}$ which is jointly typical with all the $\vec{y}_{D_i}$.  The decoder on receiving a $\vec{y}_{D_i}$, finds the index of the bin into which it falls and reports this as the message $W_i$. This is our pruned
codebook for the level-$2$ block $t_3$. Similarly pruned codebooks are
generated for all the $T_3$ level-$2$ blocks.

{\em Performance Analysis:}

As before, we first find the rate constraints for the $t_{3}$-th
level-$2$ block, so that arbitrarily small probability of error can
be achieved for decoding the message at this block. For this
purpose, we first need to establish a lower bound on the size of
$\mf{Z}_{D_{i}}$. Towards this, we have the following lemma.

\blem \label{lem:code_counting}
  \beqa
 |\mf{Z}_v| & \gto & 2^{T_2( H(\vec{Y}_v \vert F_{\mc{V}_R} = f_{\mc{V}_R}) -  N T_{1 }\kappa )} \text{ w.h.p.~as } T_2 \rightarrow \infty, \label{eq:Zv_counting}
\eeqa where $N=|\mc{V}|-1$. \elem
 We use the notation $a_n \gto 2^{nb}$, to denote the existence of a nonnegative sequence
  $\epsilon_n \rightarrow 0$ such that $a_n >
2^{n(b-\epsilon_n)}$. Notations $\lto$ and $\eqo$ are used in a similar sense
in the rest of the paper. For a sequence of events $\mf{E}(n)$
indexed by $n$, we use ``$\mf{E}(n)$ w.h.p. as $n \rightarrow
\infty$'' to denote that $ \prob \{ \mf{E}(n) \} \rightarrow 1,
\text{ as } n \rightarrow \infty.$ In this section $T_2$ will play
the role of $n$.

\bpf The details of the proof are in Appendix~\ref{app:B}. To prove
this lemma, we use the randomness in the choice of the pruned
subsets $\mf{S}_v$. \epf

The following lemma then characterizes the achievable rates,
$R_{i}(t_{3})$, by our pruned scheme in a $t_{3}$-th level-$2$
block.

\blem \label{lem:prunedMarton} As long as $R_{i}(t_{3})$
 satisfies, $\forall i=1,\ldots,J, \forall \mc{J} \subseteq \lbr 1,\ldots,J
\rbr$, \beq R_{\mc{J}}(t_3) \leq \frac{1}{T_1}
H(\vec{Y}_{D_\mc{J}}|F_{\mc{V}_{\mc{R}}}=f_{\mc{V}_{\mc{R}}}(t_3)) -
|\mc{J}| N\kappa , \label{eq:level2_perf} \eeq the probability of
error can be made arbitrarily small by choosing a large enough
$T_{2}$. \elem \bpf The result follows from the fact that, our
pruned version of the Marton scheme bins $\mf{Z}_{D_{i}}$ instead of
$\mc{T}_{\delta}^{T_{2}}(\vec{Y}_{D_{i}})$. The details of the proof
are in Appendix \ref{app:prunedMarton}. \epf

As before, we can now  average over the rates in each level-$2$
block, to get the overall rate $R_{i}$. Under the pruned scheme,
each rate is reduced by $N\kappa$. Thereby, we have the following
lemma.

\blem \label{lem:prunedDSN} A rate tuple $R = (R_{1},\ldots,R_{J})$
is achievable for the DSN, using the pruned scheme $\mc{P}(\kappa)$,
if $\forall~\mc{J}$, there is a product distribution
$\prod\limits_{v\in\mc{V}} p(X_{v})$ (denoted by $Q_{p}$) such that
\beq R_{\mc{J}}  \leq \bar{C}_{\mc{J}}(Q_{p})- |\mc{J}|N\kappa. \eeq

\elem

\subsubsection{Code for the Gaussian Network \label{sec:lifting}}

We will now show that the pruned scheme for the DSN can be lifted to
a scheme for the Gaussian network to establish the following result.

\blem \label{lem:pruned_equals_gaussian} If $\kappa = \log ( 12N-2)
+ 11$, then any rate tuple $(R_1,R_2,...,R_J)$ that can be achieved
in the DSN using the pruned coding scheme $\mc{P}(\kappa)$ can also
be achieved in the Gaussian network. \elem

\bpf Note that the pruned coding scheme $\mc{P}(\kappa)$ for the DSN
operated over three levels of nested blocks. This scheme is now
adapted to the Gaussian network as follows. We will use
$\underline{X}_v, \underline{Y}_v$ to denote the level-$2$ blocks in
the Gaussian network (corresponding to $\vec{X}_v,\vec{Y}_v$ in the
DSN).

\ben
\item The coding over level-$2$ blocks is performed in the same way as in the DSN. We will describe the operation during the $t_3$-th level-$2$ block.  In this level-$2$ block the relay mapping is fixed to $f_{\mc{V}_R} (t_3)$, we will suppress this conditioning henceforth for notational convenience.
\item Given a message $W_i(t_3)$, the source transmits the same vector that it would have transmitted in the DSN, i.e., $\underline{x}_S^{T_2} (W_1(t_3),...,W_J(t_3)) = \vec{x}_S^{T_2} (W_1(t_3),...,W_J(t_3))$.
\item Node $v$ receives $\underline{y}_v^{T_2}$ and decodes $\vec{y}_v^{T_2}$, the corresponding vector that it would have received in the DSN.
    \blem If $\kappa = \log( 12N-2) + 11$, then the probability of error for decoding  the received vector $\vec{y}_v^{T_2}$ in the DSN from the corresponding  vector $\underline{y}_v^{T_2}$ in the Gaussian network goes to zero as $T_2 \rightarrow \infty$. \elem
    \bpf There is a natural joint distribution on $(\vec{X}_S,\vec{Y}_v,\underline{Y}_v)$ given by
    \beqa p({\vec{X}_S,\vec{Y}_v,\underline{Y}_v}) & = & p({\vec{X}_S}) p({\vec{Y}_v | \vec{X}_S}) p({\underline{Y}_v | \vec{X}_S}) \\
    & = & p({\vec{X}_S}) p({\vec{Y}_v | \vec{X}_S}) p({\underline{Y}_v | \underline{X}_S}). \eeqa
    Here $p({\vec{Y}_v | \vec{X}_S})$ and  $p({\underline{Y}_v | \underline{X}_S})$ are a function of the relay mappings.
    The relay node in the Gaussian network finds the $\vec{y}^{T_2}_v \in \mf{S}_v$  such that\footnote{While the relay also knows that the received vector should be contained in $\mf{Z}_v \subseteq \mf{S}_v$,
      we do not explicitly use this information in the decoding step.  } \beqa (\underline{y}^{T_2}_v,\vec{y}^{T_2}_v) \in
      \mc{T}^{T_2}_{\delta}(\underline{Y}_v,\vec{Y}_v). \eeqa
     $\mf{S}_v$ is a random $2^{-T_{1}T_{2}\kappa}$ fraction of the typical set $ \mc{T}^{T_2}_{\delta}(\vec{Y}_v)$.
     Therefore
          \beqa |\mf{S}_v| & \eqo & 2^{-T_2 T_1 \kappa } 2^{{T_2} H(\vec{Y}_v)}. \eeqa
          If we choose $ \kappa = \log( 12N-2) + 11$, then $T_1 \kappa > H(\vec{Y}_v | \underline{Y}_v)$ (see \cite{AK10} for
          a similar argument). Therefore,
          \beqa |\mf{S}_v| & \lto & 2^{-T_2 H(\vec{Y}_v | \underline{Y}_v) } 2^{T_2 H(\vec{Y}_v) } \\
\Rightarrow |\mf{S}_v| & \lto & 2^{T_2(I(\vec{Y}_v;
\underline{Y}_v))}. \eeqa This condition ensures that the typical
set decoding succeeds with high probability, and thus we can decode
$\vec{Y}_v^{T_2}$ from $\underline{Y}_v^{T_2}$. \epf

\item Any node $v$ now performs the same mapping  as in the DSN,
\beqa \vec{x}_v^{T_2} =  {f}_v ( \vec{y}_v^{T_2}), \eeqa
 and transmits the vector $\underline{x}_v^{T_2} = \vec{x}_v^{T_2}$.
\item The destination $D_i$ reconstructs the corresponding vector $\vec{y}_{D_i}^{T_2}$ from $\underline{y}_{D_i}^{T_2}$,
and since decoding is possible in the DSN, it is possible in the Gaussian network as well.
\een

To conclude: this procedure yields a code for the Gaussian network that can
achieve the same reliable rate of communication as the pruned code in the DSN. \epf

\subsubsection{A Simplified Coding Scheme}
Our coding scheme for the deterministic and the Gaussian networks
comprised coding over three levels of blocks. We will show now that, if we consider the maximization of a linear functional of the rate, the third level is not necessary  (i.e., $T_3=1$ is sufficient) in an operational scheme, and is used only as a random coding technique.
 Suppose we want to
maximize a certain linear functional of the rate, \beqa \max_{R \in
\bar{\mc{C}}(Q_{p})} \sum_{i} \lambda_i R_i. \eeqa In
Section~\ref{sec:codingSchemeDN}, we saw that the average of the
rate across various level-$2$ blocks yields the rate $R_{i} =
\frac{1}{T_3} \sum_{t_{3}=1}^{T_{3}} R_{i}(t_3)$. Thus \beqa
\sum_{i} \lambda_i R_i = \frac{1}{T_3} \sum_{t_{3}=1}^{T_{3}}
\sum_{i}  \lambda_i R_i(t_3), \eeqa which implies that there exists
a $t_3$ such that $ \sum_{i} \lambda_i R_i(t_3) \geq \sum_{i}
\lambda_i R_i$. If we use the corresponding relay encoding functions
$f_{\mc{V}_R}(t_3)$ throughout, then we achieve the best possible
linear functional of the rate under this scheme. Thus coding over
$T_3$ level-$2$ blocks is unnecessary for maximizing a linear
functional in the rate region.
Thus coding over multiple relay transformations is only a proof
technique and not a method of operation of the network for maximizing a linear functional of the rate.

By the convexity of the rate
region of the scheme, all extreme points in the rate region can be achieved
 by setting $T_3 = 1$. To achieve any point inside the rate region, time sharing of these schemes will be required in general.
Since the relay encoding functions $f_{\mc{V}_R}(t_3)$
required to achieve this may be different for different extreme points, time sharing can be thought of as being equivalent to coding with a larger $T_3$ .

\subsection{Non-layered Networks} \label{sec:nonLayered}
Consider a general network specified by the set of vertices
$\mc{V}$. In \cite{NetCod,ADT09}, it has been shown in the context of unicast traffic that any network can be
unfolded in time to get a layered network. In the broadcast scenario here, we use the same
procedure. The layering strategy is briefly described below. Similar to \cite{ADT09}, the level-$1$ (inner) block is now over
$KT_{1}$ time symbols.
The relay node still does random mappings over blocks of $T_{1}$ symbols, however the transmit vector in $k$-th block, for $k=1,\ldots,K$, now depends on the last $k-1$ received blocks.
This relaying scheme can then be represented as a layered-network in time.
The induced layered network
then has $K$ layers, and each layer has $|\mc{V}|+J+1$ nodes - the
$|\mc{V}|$ nodes $v[k],k=1,2,...,|\mc{V}|$ corresponding to the
network at time slot $k$ and $J+1$ special nodes
$T[k],R_1[k],...,R_J[k]$ that act as virtual transmitters and
receivers that act as transmit and receive buffers, holding all the information about source message in the case of transmit buffer and holding all the received information in the case of received buffer. The set of edges connecting the adjacent layers is
derived from the network  by the following procedure: \ben \item
$v_1[k]$ is connected to $v_2[k+1]$ with a link $h_{v_1,v_2}$ which
is the channel corresponding to the link from  $v_1$ to $v_2$ in the
original network.
\item Memory inside a node is maintained by connecting $v_i[k]$ to $v_i[k+1]$ using an orthogonal and infinite capacity link.
\item The transmit buffer is maintained by connecting $T[k]$ to $T[k+1]$ and also $T[k]$ to $S[k+1]$ using
orthogonal and infinite capacity links.
\item Receive buffer for the the destination node $j$ is maintained by connecting $R_j[k]$ to $R_j[k+1]$ and from $D_j[k]$ to $D_j[k+1]$ using orthogonal and infinite capacity links.
\item A $0$-th layer with $S = T[0]$ alone, and a $K+1$-th layer with $D_j = R[k+1] , j=1,...,J$ are added to serve as the source and destinations in this unfolded network.
\een

Let us consider the cut-set bound (normalized by $K$) between the
source and the destination nodes $\mc{J}$ for this unfolded network
$\bar{C}^{K-\text{unf}}_{\mc{J}} (Q)$ and for the original network
by $\bar{C}^{\text{org}}_{\mc{J}} (Q)$. Then, we have the following
lemma (See proof of Lemma $5.1$ in \cite{ADT09}). \blem
\label{lem:adt_cuts}  $ \bar{C}^{K-\text{unf}}_{\mc{J}} (Q) =
\frac{K - |\mc{V}|}{K} \bar{C}^{\text{org}}_{\mc{J}} (Q). $ \elem

If we take $K$ large enough, the effect of the penalty term
$\frac{K-|\mc{V}|}{K}$ can be made as small as desired. Lemma~\ref{lem:adt_cuts},
combined with our result for layered networks, yields the desired
result for general (non-layered) networks.

\section{Discussion} \label{sec:discussion}

In this section, we first review linear deterministic broadcast networks, and the capacity-achieving schemes for such networks. We then show how these schemes are reciprocal to the schemes for the multi-source single-destination deterministic networks. Then we proceed to identify the intuition connecting the reciprocal schemes based on a contrast between {\em transmitter-centric} and {\em receiver-centric} viewpoints, and use this intuition to point out the inspiration for the Gaussian broadcast network scheme. At the end of this section, we point out the channel state information required at various nodes in order to implement these schemes.

\subsection{Linear Deterministic Networks (LDN)} \label{sec:LDN}

A deterministic network of particular interest is the linear
finite-field broadcast deterministic network \cite{ADT09}. The
inputs and outputs are vectors over a finite field,
i.e.~$\mc{X}_{j}=\mc{Y}_{j}=\mb{F}_{p}^{q}$, for some prime $p$ and
$q\in\mb{N}$. The channels are linear transformations over this
finite field i.e., \beq y_j[m] = \sum_{i\in N_{j}} G_{i,j} x_i[m],
\label{eq:LDNchannelmodel} \eeq where $G_{i,j} \in
\mb{F}_{p}^{q\times q}$. In particular,   $G_{i,j}$ are often assumed
to be ``shift" matrices. The linear deterministic model captures
wireless signal interaction like interference and broadcast and on
the other hand has an algebraic structure that can be exploited for
understanding schemes in this network.

\bcor For the linear deterministic broadcast network the capacity
region is in fact the cut-set bound.  \ecor
\bpf We can show that for the linear
deterministic network the cut-set bound, $\bar{C}_{\mc{J}}(Q)$, is
maximized by uniform and independent distribution of $\lbr X_{v} |
v\in \mc{V} \rbr$. Therefore \beq \bar{C}_{\mc{J}}(Q) =
\bar{C}_{\mc{J}}(Q_{p}) = \min_{\Omega \in \Lambda_{\mc{J}}}
\text{rank}(G_{\Omega,\Omega^{c}}), \eeq where
$G_{\Omega,\Omega^{c}}$ is the matrix relating the vector of all the
inputs at the nodes in $\Omega$ to the vector of all the outputs in
$\Omega^{c}$ induced by \eqref{eq:LDNchannelmodel}. The inner bound
has already been shown for the general deterministic network in
Theorem~\ref{thm:capGenDet}. This proves the corollary. \epf

An important question is whether simpler linear schemes are optimal for
these networks. It has already been shown in \cite{ADT09} that for
the single-source single-destination relay network, linear mappings
at all nodes suffice. The intuition behind the proof is that, when the relay nodes
randomly pick transformation matrices, the resulting matrix between the source and the
destination has rank equal to the min-cut rank of the network, with high probability. Therefore, if the rate is lesser than the min-cut rank, random
linear coding at all nodes (including the source but not the destination) ensures an
end-to-end full-rank matrix and the destination, knowing all these
encoding matrices, picks up a decoding matrix, which is the inverse of the end-to-end matrix. This intuition is
then used to obtain schemes in the general deterministic relay network
and the Gaussian relay network in \cite{ADT09}, where the relays
perform random mapping operations resulting in an induced end-to-end
channel between the source and the destination. Then the source uses a random code to map the messages, and the destination performs a typical set decoding.
 It has also been shown in \cite{GIZ09} and \cite{YS09}, that for the linear deterministic relay network,
 restricting the relay mappings to permutation matrices is without loss of  optimality.
 The next corollary claims a similar result even for the linear deterministic broadcast network.
\bcor
For linear deterministic broadcast network, linear coding at every node is sufficient to achieve capacity. Furthermore, the mapping at relay nodes can be restricted to permutation matrices.
\ecor
Although this can be proved
directly, we will use the connection between linear coding and
reciprocity to prove this in the next section.

\subsection{Reciprocity}

The reciprocal of a Gaussian communication network (with unit power constraint at all nodes) with multiple
unicast flows can be defined as the network where the roles of the
sources and the destinations are swapped. Note that any channel coefficient that captures the signal attenuation between a pair of nodes is the same in either direction.
For a linear deterministic network, the reciprocal network was defined in \cite{RPV09} as the network where the roles of the
sources and the destinations are swapped, and the channel matrices are chosen as transposes of each other in the forward channel for the network and its reciprocal.

While it is unresolved whether a given network and its reciprocal have the same capacity region, many interesting examples are known for which this is true. For some cases, this reciprocity is applicable even at the scheme level. \beit
\item {\em Wire-line networks} can be considered as a special case of wireless networks studied here. It has been shown in \cite{R07} that wire-line networks are reciprocal (also called reversible in the literature) under linear coding.
\item In \cite{RPV09}, it was shown that reciprocity, under linear coding, can be extended naturally to the {\em linear deterministic network}. The reciprocity was shown at the scheme level and the coding matrices at each node can be obtained from the reciprocal network.
\item In Gaussian networks, duality has been shown, \cite{VJG03, VT03}, between the multiple access channel (MAC) and broadcast channel (BC), where it was shown that the capacity region of the MAC is equal to the capacity region of the BC under the same sum power constraint. This duality was also shown, interestingly,  at the scheme level between the dirty-paper pre-coding for the BC and the successive cancellation for the MAC.
\eeit

The reciprocal network corresponding to the broadcast network
studied here is the network with many sources and one destination.

\subsubsection{Sufficiency of Linear Coding in LDN}
The multi-source single-destination network has been studied in \cite{LKEC10, Perron, KM10, BCM10}, the capacity region for the linear deterministic
network with many sources and one destination is established and it
is further shown that linear coding is sufficient to achieve this.
This is done by converting the problem to the case of single-source
single-destination by adding a super-node and connecting all the
source nodes to the super-node by orthogonal links with capacities
equal to the rate required for that source. Since random linear coding at the source and the relays works for the single-source single-destination network, it works for this network too. Therefore, the source nodes and the relay nodes perform random mappings, and the destination, knowing the source
and relay mappings, can then carefully pick the decoding matrix that
inverts this overall matrix.
Since this coding is linear, we can use the reciprocity result of
\cite{RPV09}, to show that any rate achievable in the dual multiple-source single-destination network is also achievable in the single-source multiple-destination case.
Along with the fact that the cuts are reciprocal in these two networks, this implies that linear coding is optimal even in the case of the linear deterministic broadcast network. This result has also been shown in \cite{KM10} without using reciprocity by adopting an algebraic approach.

Furthermore, from the results of
\cite{GIZ09} and \cite{YS09}, it can also be shown that the sources
and the relays can pick up specific permutation matrices for the single-source single-destination network. The above argument can then the extended to show that a coding scheme involving only permutation mappings at the relays is sufficient for the linear deterministic broadcast network.

\subsection{Receiver-Centric Vs. Transmitter-Centric Schemes: Intuition for the Gaussian Broadcast Network Scheme}
We now continue on our discussion on duality for linear deterministic network to illustrate how these ideas lead us to a scheme for the Gaussian broadcast network. We begin by defining two viewpoints in which schemes can be constructed. A {\em transmitter-centric scheme} in one in which the scheme is constructed from the viewpoint of the transmitter, where the codebook at the transmitter is first selected using a random coding argument and then the receiver chooses its de-codebook in accordance with the realization of the transmit codebook. In contrast, in a {\em receiver-centric scheme}, we fix the decode-book,
which comprises of the mappings from the received vectors to the messages, and based on these mappings,
the transmitter chooses its codebook to ensure low probability of error.

Because random coding is done at the source, we can think of this scheme as first constructing the transmitter codebook in a random manner
and the receiver then constructs its de-codebook as a function of the realization of the transmit codebook.
While the scheme for the linear deterministic multi-source
network is transmitter-centric, the scheme for the linear-deterministic broadcast network is {\em receiver-centric}.

\subsubsection{A Point-to-Point Channel}

For a point-to-point channel,
 the usual random coding scheme \cite{elemIT} can be regarded as either a transmitter-centric scheme, which is the traditional viewpoint (since the random codebook is thought of as being constructed at the source), or as a receive-centric scheme. It can be viewed as a receiver-centric scheme, because, at the receiver we construct a vector quantization codebook (alternately viewed as the decode-book) or rate $R$,
 which ``quantizes'' the received signal $y^T$ to a vector $x^T(m)$, for some $m$ where $m$
is the message index and $x^T(m)$ is the $m$-th quantization codeword. Now the source sets its codebook to be equal to the vector quantization codebook at the destination. This scheme is the same scheme as the usual random coding scheme.  The distinction between transmitter-centric and receiver-centric schemes in this example is therefore one of personal preference, rather than an enforced one.

\subsubsection{Multiple-Access Vs. Broadcast Channel}

In some networks, we may not have the luxury to use the two viewpoints simultaneously, in which case we need to choose between the two. In the capacity-achieving coding scheme for the multiple access channel \cite{elemIT}, the random coding is done at the transmitters and the receiver does joint typical-set decoding, based on the specific codebooks constructed at the sources. This provides a good example of a transmitter-centric scheme.

In contrast, for the two-user broadcast channel, we can now view the Marton coding scheme (\cite{M79,EM81}), used in Sec.~\ref{sec:sourceMap}, as a receiver-centric scheme. In this scheme, there are two auxiliary random variables, $U_1$ and $U_2$, which we view as corresponding to the vector quantization variables at the two users.
The receiver $i$ can be thought of as constructing a vector quantization codebook which ``quantizes'' the received vector $Y_i^T$ to $U_i^T(w_i)$, where $w_i$ is an index belonging to a set larger than the set of  all messages to user $i$, and bins the set of all $w_i$ to the message $m_i$ for user $i$.
The transmitter, to transmit a message pair $(m_1,m_2)$, finds a pair $(w_1,w_2)$ such that $U_1^T(w_1),U_2^T(w_2)$ is jointly typical. From this viewpoint, the receivers are choosing random de-codebooks, and the transmitters are choosing specific codebooks to be a function of the realization of the de-codebook. Thus the coding scheme can be viewed as a receiver-centric one.

\subsubsection{Multiple-Access Vs. Broadcast in Linear-Deterministic Networks}

\begin{figure}[htb]
\centering
\subfloat[Random coding scheme for multi-source single-destination network]{
\scalebox{0.5}{\input{figs/reciprocity1.pstex_t}}
\label{fig-reciprocity1}
}
\hspace{0.1in}
\subfloat[Dual random coding scheme for single-source multi-destination network]{
\scalebox{0.5}{\input{figs/reciprocity2.pstex_t}}
\label{fig-reciprocity2}
}
\caption{Reciprocity in linear deterministic networks}
\label{fig-reciprocity}
\end{figure}
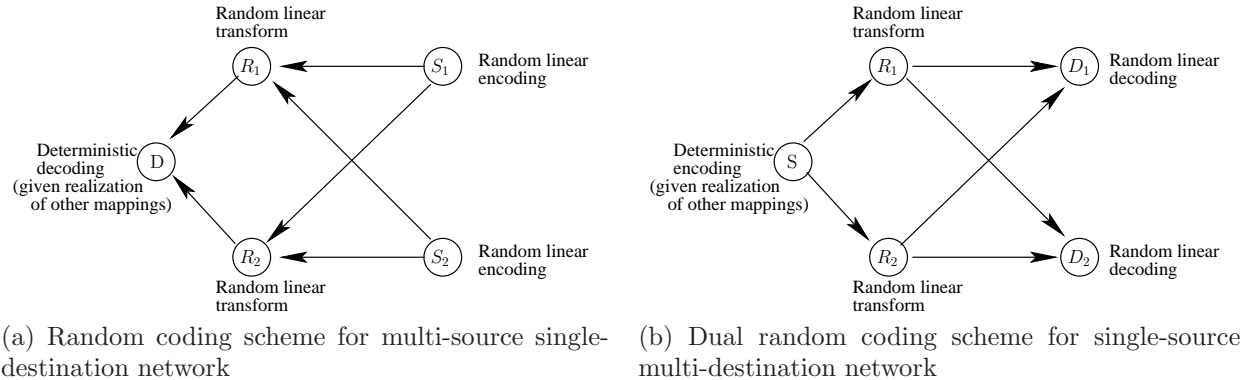

From \cite{Perron,LKEC10}, we know that a transmitter-centric scheme, where the sources and the intermediate nodes perform random coding,
 is optimal for the many-source single-destination problem in the linear determinstic setup. Intuition suggests that a natural receiver-centric method should work for the reciprocal network (i.e., single-source multiple-destination network). In particular, the relays perform random mappings, and the destinations perform ``random decoding'', i.e., they fix a random linear mapping from the received vector into a smaller message vector. Once these mappings at the relay and the destinations are fixed, the source evaluates the induced linear broadcast channel between the source and the various destinations; and constructs a linear broadcast code for this channel. This scheme can then be shown to be optimal for the broadcast network, because this is the reciprocal of the linear random coding scheme, which is optimal for the multi-source single-destination network, as shown in Fig.~\ref{fig-reciprocity}.

%
%
%\begin{figure}[htb]
%\centering
%\subfigure[Random coding scheme for multi-source single-destination network]{
%\scalebox{1}{\input{figs/reciprocity1.pstex_t}}
%\label{fig-reciprocity1}
%}
%\subfigure[Dual random coding scheme for single-source multi-destination network]{
%\scalebox{1}{\input{figs/reciprocity2.pstex_t}}
%\label{fig-reciprocity2}
%}
%\label{fig-reciprocity}
%\caption{Reciprocity in linear deterministic networks}
%\end{figure}
%
%
%
%

\subsubsection{Scheme for Gaussian Broadcast Networks: Lifting Scheme as a Receiver-Centric Scheme}

The general idea for the scheme for the linear deterministic broadcast network is the foundation of our scheme for the Gaussian broadcast
network in Sec.~\ref{sec:Gaussian}. In order to build the scheme for the Gaussian network, we first construct a scheme for general deterministic networks (of which the DSN is a special case) and then lift the scheme from the DSN to the Gaussian network. In case of the linear-deterministic broadcast network, the source-mapping depended on the specific relay transformations used, not just on the probability distribution used to create the relay transformation. Extending this idea, we would like to construct a scheme for the general deterministic network, where the source codebook is a function of the {\em specific relay transformation}. Indeed, we resolve this problem by constructing a Marton scheme at the source for the vector broadcast channel induced by the specific relay mappings.

Next, the scheme for lifting codes from the DSN to Gaussian relay networks proposed in \cite{AK10}
 requires each node, including the destination, to prune their received vectors to a restricted set to ensure that the received vector in the DSN
 can be decoded from the received vector in the Gaussian network. Since this scheme restricts the received codewords at the destination,
 this scheme also naturally fits into a receiver-centric viewpoint.

 While the lifting procedure proposed there works only for single-source single-destination networks, we extend the procedure to our specific scheme for broadcast network.  We achieve this by designing a pruned Marton code, in which the receivers are guaranteed to receive vectors which are in the pruned set. Instead of binning the set of all possible received vectors into messages, as we would for a broadcast channel, we now bin only the pruned received vectors to construct the pruned Marton coding scheme. The natural alignment of the receiver-centric viewpoints of the Marton scheme and the lifting scheme allows us to construct the scheme for the Gaussian broadcast network.

\subsection{Approximate Reciprocity in Gaussian Multi-Source and Broadcast Networks}

In this section, we will demonstrate that there is an approximate reciprocity in the capacity regions of a Gaussian multi-source network and the corresponding reciprocal Gaussian broadcast network.

In our model, we have assumed, without loss of generality, the average transmit power constraint of unity at each node. We
have also assumed that the reciprocal network, in addition to having the
same channel coefficients, also has unit power constraints at each
node. However, it is  not clear if this is the ``right'' way of defining
the corresponding reciprocal network. For instance, in \cite{VJG03,VT03}, MAC-BC duality was shown under the assumption of
same total transmit power in both networks; however this power could be divided amongst the nodes in a different manner in the forward and reciprocal networks.
 Under this assumption, it was shown that the capacity region of the two
networks was identical. However, since we are concerned only about approximate reciprocity in this section, which is a weaker form of reciprocity, our definition of
unit power constraint everywhere will be sufficient to show approximate reciprocity.

In \cite{Perron} and \cite{LKEC10}, a coding scheme is given for the
Gaussian network with many sources and is shown to achieve the
cut-set bound region within a constant gap, which depends only on
the network gain. In Sec.~\ref{sec:Gaussian}, we have showed that for the Gaussian broadcast
network also, we can achieve the cut-set bound region within a
constant gap. As a result, to show that the capacity region of
the two networks are themselves within a constant gap, which depends
only on the network topology and not on the channel gains, all we need to do is to observe that cut-sets of the reciprocal networks are within a constant gap of each other.
Note that the cut-set bounds corresponds to MIMO point-to-point channel where all the nodes on the source side
of the nodes can be thought of as transmit antennas and all the
nodes on the destination side can be thought of as receive antennas.
The relationship then between a cut in a network and the
corresponding cut in the reciprocal network is the same as the
relationship between a MIMO channel with channel matrix $H$ and the
reciprocal MIMO channel with the channel matrix $H^{T}$.  The
reciprocity of MIMO channel has been shown in \cite{Telatar}, under
equal total transmit power, i.e.~the capacity of the two networks is
the same. It can be further shown that restricting to per node power
constraint only leads to a loss which does not depend on the channel
gains. Therefore,  we can show that the cut-set bounds are
reciprocal.

\subsection{Induced Coordination in Relays' Transmission}

\begin{figure}[htb]
\begin{center}
\scalebox{0.65}{\input{figs/coordination.pstex_t}}
\end{center}
\caption{MISO broadcast channel as a special case of broadcast network} \label{fig-coordination}
\end{figure}
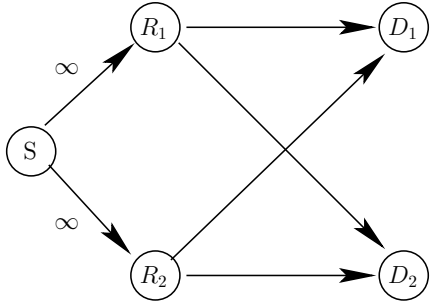

Let us consider a simple example for a broadcast relay network comprised of a single source, two relays and two destinations, shown in Fig.~\ref{fig-coordination}. The link between the source to the two relays is infinite, which implies therefore that the network is essentially a MISO broadcast channel with two transmit antennas and two receivers, each with a single antenna. It is clear that for a MISO broadcast channel, independent coding across the two relays is insufficient to even obtain the best possible degrees-of-freedom. Therefore, any scheme that is approximately optimal needs to perform coordinated transmission at the relays.

In the proposed scheme, the relays perform quantization followed by independent encoding of the quantized bits into transmitted vectors. At a first glance, a scheme in which the relays are performing independent mappings seems incapable of attaining good performance because of the inability to induce coordination. However two key features in the proposed scheme help avoid this pitfall. \begin{itemize}
\item The relays $R_1$ and $R_2$ perform quantize-and-encode relaying in the aforementioned example, in spite of the fact that they can decode the source message completely. Had the relays decoded the source message and performed independent encoding, there is no possibility of achieving the degrees-of-freedom of even this simple broadcast network.
\item The source takes into account the specific realizations of the relay mappings and constructs the coding scheme. This ensures that from the point of view of the receiver, the signals transmitted by the two relays appear coordinated. In particular, in this example, since the channel from $S$ to $R_i$ is infinitely good, the relay $R_i$ quantizes the received signal to a very fine degree and encodes this for transmission to the destination. This gives the source many degrees-of-freedom to encode information in the various least-significant-bits of its transmission, so that after the relay mappings, the relay transmissions appear coordinated.
\end{itemize}

\subsection{Channel State Information}
We now examine the channel state information required at the various nodes for the schemes proposed in Sec.~\ref{sec:deterministic} and Sec.~\ref{sec:Gaussian} for deterministic and gaussian broadcast networks.

\subsubsection{Deterministic Broadcast Networks}
For deterministic networks, the following channel state information
is required: \ben
\item All non-source non-destination nodes are unaware of any channel state information.
\item We assume that each destination knows the distribution of the received vector $\vec{Y}_{D_i}$, and the transmitted rate $R_i(t_3)$ for each $t_3$. The destination  bins the set of all typical vectors into $2^{T_1 T_2 R_i}$ bins corresponding to the messages, and uses this as the decoding rule.
\item The transmitter is assumed to have full CSI, and knows the relay mappings at all nodes and also the binning scheme at the destinations. The transmitter construct the codebook using the same binning scheme as the receiver.
\een

Thus this scheme has the interesting property that if the
transmitter had all knowledge, the intermediate nodes have zero
knowledge and the destination has a little knowledge (about the distribution of the received vector), then the
same rate can be achieved as the complete CSI case.

This is dual to the situation of the multi-source single-destination network, where
the receiver having full knowledge, intermediate nodes having zero
knowledge and the transmitters having a little knowledge (about the distribution of the transmitted vector) can
achieve the same rate as full channel knowledge.

\subsubsection{Gaussian Broadcast Network}

For Gaussian networks, the following channel state information is
required: \ben
\item All non-source non-destination nodes are unaware of any channel state information. Node $v$ however knows the probability distribution of the {\em received {and} transmitted vectors} $p_{Y_v},p_{X_v}$ for the corresponding DSN, which will be used to calculate the relay mappings. The node $v$ also needs to use the received vector distribution to pick a pruned subset of the typically received vectors in the corresponding DSN.
\item We assume that each destination knows the distribution of the received vector in the corresponding DSN $\vec{Y}_{D_i}$, and the transmitted rate $R_i(t_3)$ for each $t_3$. The destination maps bins the set of all typical vectors into $2^{T_1 T_2 R_i}$ bins corresponding to the messages, and uses this as the decoding rule.
\item The transmitter is assumed to have full CSI, and knows the mappings used at all the nodes and also the binning scheme at the destinations. The transmitter then uses the same binning scheme used at the receiver.
\een

This scheme has the interesting property that if the
transmitter had all knowledge, the intermediate nodes and the
destination have some knowledge, then the same rate can be achieved as the
complete CSI case.

\section{Generalizations} \label{sec:generalization}
 In this section, we present various generalizations of our result, for half-duplex networks in Section~\ref{sec:half-duplex}, for networks with multiple antenna in Section~\ref{sec:MIMO} and for broadcast wireless networks, where some set of nodes demand the same information and other nodes demand independent information in Section~\ref{sec:bc-mc}.

\subsection{Half Duplex Networks} \label{sec:half-duplex}
Our discussion so far has been restricted to the context of full duplex scenario.
A network is said to be {\em half duplex} if the nodes in the network
can either transmit or receive information, but not do both  simultaneously. Therefore
the network needs to be {\em scheduled} by specifying which nodes
are listening and which nodes are transmitting at any given time
instant. Let the set of all possible half-duplex schedules at any
time instant be $\mc{H}$. An edge $e_{ij}$ is said to be {\em
active} at time slot $k$ if $v_i$ is transmitting and $v_j$ is
receiving at that time slot.

Consider $K$ time slots and at any time instant $k$, let $h_k \in
\mc{H}$ be the half duplex schedule used, and $h^K$ be the sequence $h_1,h_2,...,h_K$. We consider only {\em static schedules} here, that is,
schedules that are specified apriori and do not vary depending on
dynamic parameters like channel noise. For any static schedule
$h^K$, we can unfold the network graph with respect to that
schedule. This procedure is performed in \cite{ADT09}, and is the
same as the procedure in Section~\ref{sec:nonLayered}, except for
the following difference: $v_1[k]$ is connected to $v_2[k+1]$ with a
link $h_{v_1,v_2}$ only when $e_{v_1v_2}$ is active at time slot
$k$.

Given that the network is operated under a schedule $h^K$, we define the set of all rate pairs achievable as the {\em capacity region under the schedule} $h^K$.
An upper bound on the capacity region under the schedule $h^K$ is given by the cut-set bound in the unfolded layered network corresponding to the schedule. This rate can be achieved within a constant gap by using Theorem~\ref{thm:main}. Thus for any schedule $h^K$, any rate tuple within the constant $k = O(|\mc{V}| \log (|\mc{V}|))$ of the cut-set bound  can be achieved (to within a constant number of bits) using that schedule and then using the scheme of Theorem~\ref{thm:main} for the unfolded layered network. Now, we can optimize over all schedules $h^K \in \mc{H}$ allowed under the half-duplex constraints. Thus, the capacity region of the network under static half-duplex scheduling is the union over all possible schedules of the capacity region under schedule $h^K$. Therefore, any rate tuple $(R_1,...,R_J)$ such that $(R_1 + k ,R_2 + k,...,R_J +k)$ is in the capacity region of the network under static half-duplex scheduling can be achieved by using the method described here.

\subsection{MIMO} \label{sec:MIMO}

In this section, we  consider the implication of having multiple
antenna elements at each of the nodes in the network. Suppose $v$ possesses $m_v$
antenna elements, which are used for both transmission and reception. The basic result for multi-antenna broadcast networks is
the following.

\bthm \label{thm:MIMO} For the multi-antenna broadcast network, a rate vector $(R_1,\ldots ,R_J)$ is
achievable, \beq (R_1+k,\ldots,R_J+k)
\in \bar{\mc{C}} \eeq

for some constant $k$, which depends only on the number of nodes,
and not on the channel coefficients, and  $k = O(M \log M)$ where $M =\sum_{v \in \mc{V}} m_v$ . \ethm

\bpf The proof is essentially the same as the one for the single
antenna case in Section~\ref{sec:Gaussian}. We only outline
the proof below, highlighting the key distinctions.

\ben \item For a given multi-antenna Gaussian network, we first
obtain a multi-antenna DSN such that the cut-set bound for the
Gaussian network and the cut-set bound under the product form
distribution for the DSN differ only by a constant (in the same manner as we obtained Lemma \ref{lem:product_form} in Sec.~\ref{sec:DSN}. This is because
\eqref{eq:prod_form} continues to hold for multi-antenna
networks, with $|\mc{V}|$ now replaced by the total number of antennas $M$.
\item Then we use Theorem~\ref{thm:capGenDet} to show that the cut-set bound under product form distribution is achievable for the DSN.
\item We can now prune the codebook for the DSN by a factor $\kappa$, to get a pruned scheme for the DSN, $\mc{P}_{\kappa}$. The pruned scheme achieves a rate $M \kappa$ lesser than the original rate.
    \item Under the pruned scheme, the received vector in the DSN can be decoded from the received vector in the Gaussian network.
\item Therefore the DSN coding scheme can be emulated in the Gaussian network, and achieves a rate a constant $k = O(M \log M)$ lesser than the cut-set bound in the Gaussian network.
\een \epf

\subsection{Broadcast-cum-Multicast} \label{sec:bc-mc}
The broadcast network comprised of a single source $S$ and
destinations $D_1,D_2,...,D_J$ demanding independent messages at
rates $R_1,R_2,...,R_J$. Suppose that in addition there are also
other multicast destinations $M_1,M_2,...,M_L$ that demand all the
messages transmitted by the source. We call such a network a
broadcast-cum-multicast network. In this section, we will show that
even for such networks, the cut-set bound is achievable to within a
constant number of bits. This network is a generalization of both the multicast network considered in \cite{ADT09} and the broadcast network considered in the previous sections.

First we note that the cut-set bound for the broadcast-cum-multicast
network is given by the cut-set bound for the broadcast network,
along with the cut-set constraints for each multicast receiver. In
particular for the Gaussian broadcast-cum-multicast network, if
$(R_1,...,R_J)$ is achievable, then $\forall~\mc{J}\subseteq[J]$ there exists a joint distribution $Q$ such that,
\beqa R_{\mc{J}} \leq \min_{\Omega \in \Lambda_{\mc{J}}} I (X_{\Omega}; Y_{\Omega^c} | X_{\Omega^c}), \eeqa and in addition, the sum rate is constrained by all the multicast destinations since all these destinations demand all the messages transmitted by the source \beqa R_{[J]} & \leq & \min_{i
\in [L]}  \min_{\Omega \in \Lambda_{M_i}} I (X_{\Omega}; Y_{\Omega^c} | X_{\Omega^c}).
\eeqa
The set of all rate tuples inside the cut-set bound is then denoted by $\bar{C}$.

The main result for the wireless broadcast-cum-multicast network is that any rate a constant away from cut-set bound is achievable.

\bthm \label{thm:bc_mc_gaussian} For the Gaussian
broadcast-cum-multicast network, there exists a constant $k$, which
does not depend on the channel coefficients and is  $O(|\mc{V}| \log
|\mc{V}|)$, such that a rate vector of $(R_1,...,R_L)$ is
achievable whenever
\beqa (R_1+k,...,R_J+k) \in \bar{C} .\eeqa \ethm

To prove this result, we follow an approach similar to the one we took for broadcast
networks. First we will prove a result for deterministic
broadcast-cum-multicast networks. Second, we show that the Gaussian
network can emulate the deterministic superposition network with a
constant rate loss. These two steps are completed in the rest of this section.

\subsubsection{Deterministic Broadcast-cum-Multicast Network}

The next Lemma shows that for the deterministic
broadcast-cum-multicast network, the cut-set bound evaluated under
product form distributions is achievable.

\blem \label{lem:bc_mc_det} For the deterministic
broadcast-cum-multicast network, the cut-set bound under product-form distributions is achievable, i.e., a rate vector $(R_1,...,R_K)$ is
achievable if for every $\mc{J}\in[J]$ there is some product
probability distribution $Q_p$, such that,  \beqa R_{\mc{J}} &\leq&
\min_{\Omega \in \Lambda_{\mc{J}}} I \lp X_{{\Omega}}; Y_{\Omega^{c}} | X_{\Omega^{c}} \rp \\
&=& \min_{\Omega \in \Lambda_{\mc{J}}} H \lp Y_{\Omega^{c}} | X_{\Omega^{c}} \rp, \ \  \text{\em and }\\
 R_{[J]} & \leq & \min_{i \in [L]}
\min_{\Omega \in \Lambda_{M_i}} I \lp  X_{{\Omega}};Y_{\Omega^{c}} |
X_{\Omega^{c}} \rp \\
&=& \min_{i \in [L]} \min_{\Omega \in \Lambda_{M_i}} H \lp Y_{\Omega^{c}} |
X_{\Omega^{c}} \rp. \eeqa \elem

\bpf

{\em Coding Scheme:} The source operation, relaying operations and
decoding operations at the broadcast destinations remain the same as
in Section~\ref{sec:codingSchemeDN}. In addition, we need to specify
the decoding operation at the multicast destinations.

At each level-$3$ block, each multicast destination performs a
typical set decoding with the set of all possible $\vec{X}_s^{T_2}$
(corresponding to all possible messages) and finds the unique
$W_1(t_3),...,W_J(t_3)$ for which the
$(\vec{X}_s^{T_2}(W_1(t_3),...,W_J(t_3)), \vec{Y}_{M_i}^{T_{2}}) \in
T_{\delta}^{T_2}(\vec{X}_s,\vec{Y}_{M_i})$.

{\em Performance Analysis:} If the cut-set bound for the broadcast destinations is satisfied,
then the probability of error at these destinations is guaranteed to
be small (by the same analysis in Sec~\ref{sec:perfAnalDN}). We
need to analyze the error events at all multicast destinations.
The probability of error at destination $M_i$ in the $t_3$-th level-$2$ block goes to zero, with $T_2\rightarrow\infty$ if
\beqa R_{[J]}(t_3) & < & \frac{1}{T_1} I (\vec{X}_s; \vec{Y}_{M_i} | F_{r} = f_r(t_3) ) \\
 & = & \frac{1}{T_1} H(\vec{Y}_{M_i} | F_{r} = f_r(t_3)).
\eeqa

As before, since the overall rate is given by averaging the rate across all $T_3$ blocks, as $T_3 \rightarrow \infty$, the probability of error at destination $M_i$ is small if,
\beq R_{[J]} < \frac{1}{T_1} H(\vec{Y}_{M_i} | F_{r}). \eeq
Using Lemma~\ref{lem:mI}, we can relate the entropy term above to the cut-set w.r.t.~this destination, i.e for any arbitrary $\epsilon > 0,~\exists T_1$, s.t., we have \beqa \frac{1}{T_1} H(\vec{Y}_{M_i}| F_{r}) \geq \min_{\Omega \in \Lambda_{M_i}} H \lp Y_{\Omega^{c}} |
X_{\Omega^{c}} \rp -
\epsilon. \eeqa

\epf

\bcor For the linear deterministic broadcast-cum-multicast network, the cut-set bound is achieved. For the deterministic broadcast-cum-multicast channel (a deterministic broadcast-cum-multicast network in the absence of relays and destination cooperation) the cut-set bound is achieved. \ecor
\bpf In the former case, the cut-set bound under product form distribution is the same as the cut-set bound under general distributions since there is only one transmitting node in the network. The latter case can be proved by showing the cut-set bound for linear deterministic networks is optimized by product form distributions. \epf

\subsubsection{Gaussian Broadcast-cum-Multicast Network}
We will now prune and lift the coding scheme from the DSN broadcast-cum-multicast network to the Gaussian broadcast-cum-multicast network. Since the encoding at the source and relay mappings  are the same as a network with the broadcast destinations alone, the same procedure used for pruning and lifting the broadcast code in Sec.~\ref{sec:pruned_scheme} and Sec.~\ref{sec:lifting} can be used to lift the code for the DSN broadcast-cum-multicast network to the Gaussian broadcast-cum-multicast network. This procedure ensures that all nodes in the Gaussian network can decode the corresponding received vector in the DSN. Therefore, the destinations carry out the same decoding operation that they perform in the DSN. The performance analysis of the scheme is similar to the performance analysis for the broadcast network in Sec.~\ref{sec:Gaussian}, and it can be shown that any rate tuple $(R_1,..,R_J)$ that satisfies $(R_1+k,...,R_J+k) \in \bar{C}$ can be achieved, where $k = O( |\mc{V}| \log (|\mc{V}|) )$.

\appendix
\section{ Proof of Lemma~\ref{lem:mI} \label{app:A}}

\bpf Fix any $\mc{J}\subseteq\lbr 1,\ldots,J\rbr$. To prove the
lemma, we consider a communication scenario where the source needs
to send a message to a single destination which has access to
$Y_{D_{\mc{J}}}$ with rate $\tilde{R}_{(\mc{J})}$. Define, \beq
\tilde{W}_{(\mc{J})} \sim~\textrm{Uniform}~\lbr
[2^{\tilde{R}_{(\mc{J})}}] \rbr, \eeq and the mapping at the source
node, \beq F_{(\mc{J})} : \tilde{W}_{(\mc{J})} \rightarrow
\mc{X}_{S}^{T} , \textrm{which is generated using i.i.d.~}p(X_{S}).
\eeq Note that $F_{(\mc{J})}$ denotes a random source code-book and
$F_{\mc{V}_{R}}$ denotes random relay mappings. The probability of
error conditioned on a given source code-book and relay mapping is
given by, \beq \prob \lbr \mc{E} |  F_{(\mc{J})}, F_{\mc{V}_{R}}
\rbr = \prob \lbr Y_{D_{\mc{J}}}^T(\tilde{W}_{\mc{J}}) =
Y_{D_{\mc{J}}}^T(w^{\prime}) | w^{\prime} \neq \tilde{W}_{\mc{J}},
F_{(\mc{J})} , F_{\mc{V}_{R}} \rbr, \eeq and the average probability
of error, averaged across all code-books and random-relay mappings,
is given by, \beq \mc{P}_{e} \df \mathbb{E} \Lbr  \prob \lbr  \mc{E}
|  F_{(\mc{J})}, F_{\mc{V}_{R}}  \rbr \Rbr. \eeq By the coding
theorem in \cite{ADT09}, we have \beq \mc{P}_{e} \rightarrow 0
,~\textrm{as}~T\rightarrow \infty,~\forall~\tilde{R}_{(\mc{J})} <
\bar{C}_{\mc{J}}(Q_p). \eeq

Now, \beqan
I(\tilde{W}_{(\mc{J})};Y_{D_\mc{J}}^{T}|F_{(\mc{J})}, F_{\mc{V}_{R}}) &=& H(\tilde{W}_{(\mc{J})}) - H(\tilde{W}_{(\mc{J})}|Y_{D_\mc{J}}^T, F_{(\mc{J})}, F_{\mc{V}_{R}}) \\
& = & T\tilde{R}_{(\mc{J})} - \mathbb{E} \Lbr H(\tilde{W}_{\mc{J}}) - H(\tilde{W}_{(\mc{J})}|Y_{D_\mc{J}}^T, F_{(\mc{J})} = f_{(\mc{J})}, F_{\mc{V}_{R}} = f_{\mc{V}_{R}}) \Rbr \\
& \stackrel{\textrm{Fano}}{\geq}& T\tilde{R}_{(\mc{J})} - \mathbb{E} \Lbr 1 + \prob \lbr  \mc{E} |  F_{(\mc{J})}, F_{\mc{V}_{R}}  \rbr \tilde{R}_{(\mc{J})} T \Rbr \\
& = & T\tilde{R}_{(\mc{J})} - (1 + \mc{P}_e \tilde{R}_{(\mc{J})} T).
\eeqan Letting $\tilde{R}_{(\mc{J})} = \bar{C}_{\mc{J}}(Q_p) -
\epsilon_1$ and for large enough $T$, we have \beq
I(\tilde{W}_{(\mc{J})};Y_{D_\mc{J}}^{T}|F_{(\mc{J})},
F_{\mc{V}_{R}}) \geq T(\bar{C}_{\mc{J}}(Q_p) - \epsilon). \eeq
Further \beqan
I(\tilde{W}_{(\mc{J})};Y_{D_\mc{J}}^{T}|F_{(\mc{J})}, F_{\mc{V}_{R}}) &=& H(Y_{D_\mc{J}}^{T}|F_{(\mc{J})}, F_{\mc{V}_{R}}) - H(Y_{D_\mc{J}}^{T}|F_{(\mc{J})}, F_{\mc{V}_{R}},\tilde{W}_{\mc{J}}) \\
& \leq & H(Y_{D_\mc{J}}^{T}| F_{\mc{V}_{R}}) - H(Y_{D_\mc{J}}^{T}|F_{(\mc{J})}, F_{\mc{V}_{R}},\tilde{W}_{\mc{J}},X_S^T) \\
& = & I(X_S^T;Y_{D_\mc{J}}^{T}|F_{\mc{V}_{R}}). \eeqan Therefore,
\beq I(X_S^T;Y_{D_\mc{J}}^{T}|F_{\mc{V}_{R}}) \geq
T(\bar{C}_{\mc{J}}(Q_p) - \epsilon). \eeq  Since the channel is
deterministic \beqan
H(Y_{D_\mc{J}}^{T}|F_{\mc{V}_{R}}) & = & I(X_S^T;Y_{D_\mc{J}}^{T}|F_{\mc{V}_{R}}) \nn\\
& \geq & T(\bar{C}_{\mc{J}}(Q_p) - \epsilon). \eeqan The proof is
completed by choosing $T_1$ to be the maximum $T$ over all
$\mc{J}$. \epf

\section{Proof of Lemma~\ref{lem:code_counting}  \label{app:B}}
\bpf

Throughout this section, we will assume that the relay
mappings are fixed to $f_{\mc{V}_{\mc{R}}}(t_3)$ without making this explicit
in the conditioning expressions.
We will assume $N=2$ to begin with and establish a
lower bound on  $|\mf{Z}_1|$ (w.l.o.g.).
\beqa \mf{Z}_1 & = & { \{\vec{y}_1^{T_2} \in \mf{S}_1: \exists\vec{y}_2^{T_2} \in
\mf{S}_2, (\vec{y}_1^{T_2},\vec{y}_{2}^{T_2}) \in
\mc{T}_{\delta}^{T_{2}} \}} \nn . \eeqa
We will consider two cases.

{\em Case $1$:} $ \ H(\vec{Y}_2 | \vec{Y}_1) > T_1 \kappa$

Fix a $\vec{y}_1^{T_2} \in \mf{S}_1$. The set of sequences
$\vec{y}^{T_2}_2$ which are jointly typical with $\vec{y}_1^{T_2}$
is given by $\mc{T}_{\delta}^{T_{2}}(\vec{Y}_{2} \vert
\vec{y}_1^{T_2})$, and its size is of the order $2^{T_2
H(\vec{Y}_2\vert Y_{1})}$. Now $\mf{S}_{2}$ is a random $2^{-T_{2}
T_{1} \kappa}$ fraction of the typical set
$\mc{T}_{\delta}^{T_{2}}(\vec{Y}_2)$. Therefore we can show that ,
if $H(\vec{Y}_2 | \vec{Y}_1) > T_1 \kappa$, then w.h.p.~as $T_{2}
\rightarrow \infty$, \beq \mf{S}_{1} \cap
\mc{T}_{\delta}^{T_{2}}(\vec{Y}_{2} \vert \vec{y}_1^{T_2}) \neq
\emptyset,
 \eeq
i.e.~every $\vec{y}_1^{T_2} \in \mf{S}_1$ will be jointly typical
with some $\vec{y}_2^{T_2} \in \mf{S}_2$. Therefore the size of the
set is $|\mf{Z}_1| \eqo |\mf{S}_1| \eqo 2^{T_2(H(\vec{Y}_1) - T_1
\kappa )}$ w.h.p.~as $T_2 \rightarrow \infty$. Clearly \beq
|\mf{Z}_{1}| \gto  2^{{T_2} (H(\vec{Y}_1) - 2 T_1 \kappa)}. \eeq

{\em Case $2$:} $ \ H(\vec{Y}_2 | \vec{Y}_1) \leq T_1 \kappa$

Fix a $\vec{y}_1^{T_2} \in \mf{S}_1$. For $T_2$ large enough, the
probability that there exists a sequence $\vec{y}^{T_2}_2 \in
\mf{S}_2$ which is conditionally typical given any $\vec{y} _1^{T_2}
\in \mf{S}_{1}$ is given by $p \eqo 2^{T_2 ( H(\vec{Y}_2 |\vec{Y}_1)
- T_1 \kappa) }$. Consider an arbitrary subset $\mf{S}_{11}
\subseteq \mf{S}_1$ of size $2^{T_2 \Delta}$. The probability that
there is an element in $\mf{S}_{11}$ which is jointly typical with
an element in $\mf{S}_2$ is given by:
\beqa \prob \{\exists~\vec{y}_1^{T_2} \in \mf{S}_{11}: (\vec{y}_1^{T_2},\vec{y}_{2}^{T_2}) &\in& \mc{T}_{\delta}^{T_{2}}(\vec{Y}_{1},\vec{Y}_{2}) \text{ for some } \vec{y}_2^{T_2} \in \mf{S}_2 \}  \nn \\
& = & 1 - (1-p)^{2^{T_2 \Delta }}\\
& \geq & 1 - e^{- p 2^{T_2 \Delta }} \\
& = & 1 - e^{- 2^{T_2 ( H(\vec{Y}_2 |\vec{Y}_1) - T_1 \kappa +  \Delta) }} \\
& \rightarrow & 1, \text{ if } \Delta > T_1 \kappa- H(\vec{Y}_2
|\vec{Y}_1), \eeqa as $T_2 \rightarrow \infty$.

So we will set $\Delta = T_1 \kappa- H(\vec{Y}_2 |\vec{Y}_1) +
\epsilon_1$.  Thus if we divide $\mf{S}_{1}$ into disjoint sets
$\mf{S}_{1i}$ each of size $2^{T_2 \Delta}$, then each will have at
least one element in $\mf{Z}_1$. Therefore
\beqa |\mf{Z}_{1}|& \geq & \frac{|\mf{S}_1|}{2^{{T_2} \Delta}}  \\
&\eqo& \frac{2^{{T_2} (H(\vec{Y}_1) - T_1 \kappa )}}{2^{{T_2} \Delta}}  \\
 & = & 2^{{T_2} (H(\vec{Y}_1) - T_1 \kappa - T_1 \kappa + H(\vec{Y}_2 |\vec{Y}_1) ) }.   \eeqa
Therefore \beq  |\mf{Z}_{1} | \gto  2^{{T_2} (H(\vec{Y}_1) - 2 T_1
\kappa)}. \eeq
This completes the proof for the case when $N=2$. By iterating this
calculation, we can show that for a general $N$, w.h.p. as $T_2
\rightarrow \infty$, \beqa |\mf{Z}_1| & \gto & 2^{{T_2}
(H(\vec{Y}_1) - NT_1 \kappa ) }. \eeqa

\epf

\section{Proof of Lemma \ref{lem:prunedMarton} \label{app:prunedMarton}}

\bpf Let us consider the case where there are only two receivers
i.e.~$J=2$. The proof extends similarly for the general case.
$\mf{Z}_{D_i}$ is the set of all typically received codewords at the
destination $D_i$. We know from Lemma~\ref{lem:code_counting} that
$|\mf{Z}_{D_i}| \gto {2^{-T_2T_1 \kappa N}} 2^{T_2 H( \vec{Y}_{D_i}
| F_{\mc{V}_{\mc{R}}}=f_{\mc{V}_{\mc{R}}}(t_3))}$ w.h.p. as $T_2
\rightarrow \infty$.

Since we are binning the set of all $\vec{y}_{D_i}^{T_2} \in
\mf{Z}_{D_i}$  into $2^{T_2 T_1 R_i}$ bins, to ensure each bin has
at the least one codeword, we need for some $\epsilon > 0$, \beqan
2^{T_2 T_1 (R_i(t_{3}) + \epsilon)} & \leq & |\mf{Z}_{D_i}|. \eeqan
It is sufficient to have,
\beqan 2^{T_2 T_1 (R_i(t_{3}) + \epsilon)} & \leq & {2^{-T_2 T_1 \kappa N}} 2^{T_2 H(\vec{Y}_{D_i} | F_{\mc{V}_{\mc{R}}}=f_{\mc{V}_{\mc{R}}}(t_3))}  \\
\Rightarrow \ R_i(t_{3}) & < & \frac{1}{T_1}  H( \vec{Y}_{D_i} |
F_{\mc{V}_{\mc{R}}}=f_{\mc{V}_{\mc{R}}}(t_3) ) - N\kappa . \eeqan

Now, there is no error if corresponding to each message pair
$(W_1(t_{3}),W_2(t_{3}))$, the source can find a jointly typical
$(\vec{y}_{D_1}^{T_2},\vec{y}_{D_2}^{T_2})$ in the bin corresponding
to $(W_1(t_{3}),W_2(t_{3}))$. This can be done w.h.p. as $T_2
\rightarrow \infty$,  as long as there are at the least $2^{T_2
(I(\vec{Y}_{D_1};\vec{Y}_{D_2}|
F_{\mc{V}_{\mc{R}}}=f_{\mc{V}_{\mc{R}}}(t_3))+ \epsilon)}$ pairs of
$(\vec{y}_{D_1}^{T_2},\vec{y}_{D_2}^{T_2})$ in this bin. This
condition translates to \beqan
\frac{|\mf{Z}_{D_1}|}{2^{T_2T_1R_1(t_{3})}}
\frac{|\mf{Z}_{D_2}|}{2^{T_2 T_1 R_2(t_{3})}} & \geq & 2^{{T_2
(I(\vec{Y}_{D_1};\vec{Y}_{D_2}|
F_{\mc{V}_{\mc{R}}}=f_{\mc{V}_{\mc{R}}}(t_3))+\epsilon)}}. \eeqan
This is satisfied if, \beqan
 \frac{2^{T_2 \lbr H(\vec{Y}_{D_1}| F_{\mc{V}_{\mc{R}}}=f_{\mc{V}_{\mc{R}}}(t_3))) + H(\vec{Y}_{D_2}| F_{\mc{V}_{\mc{R}}}=f_{\mc{V}_{\mc{R}}}(t_3))\rbr}}{2^{T_2 T_1 (R_1(t_{3}) + R_2(t_{3}))} 2^{2 T_2 T_1 N \kappa}} & \geq & 2^{{T_2 (I(\vec{Y}_{D_1};\vec{Y}_{D_2} | F_{\mc{V}_{\mc{R}}}=f_{\mc{V}_{\mc{R}}}(t_3) ) + \epsilon)}} \\
\Rightarrow R_1(t_{3}) + R_2(t_{3})  & < & \frac{1}{T_1}
H(\vec{Y}_{D_1},\vec{Y}_{D_2}|
F_{\mc{V}_{\mc{R}}}=f_{\mc{V}_{\mc{R}}}(t_3) ) -2 T_2 T_1 N \kappa .
\eeqan

\epf

\section{Proof of Lemma \ref{lem:product_form}} \label{app:DSNapprox}

First of all, we note that the for the Gaussian network, the cut-set bound is given by letting the inputs be jointly Gaussian, i.e.~$Q=\mc{CN}\lp0,K\rp$ with $K_{jj} \leq 1$ such that

\beqa \bar{\mc{C}}^{\text{Gauss}} & \df & \lbr (R_1,...,R_J): R_{\mc{J}} \leq \bar{C}_{\mc{J}}^{\text{Gauss}} (Q) = \min_{\Omega \in \Lambda_{\mc{J}}} \log \vert  I + H_{\Omega,\Omega^{c}} K_{\Omega} H_{\Omega,\Omega^{c}}^{*} \vert,~\forall \mc{J} \subseteq[J] \rbr, \label{eq:cut-setGauss1} \eeqa
where
$H_{\Omega,\Omega^{c}}$ is defined as the matrix such that \beq
Y_{\Omega^{c}} = H_{\Omega,\Omega^{c}} X_{\Omega} +
H_{\Omega^{c},\Omega^{c}} X_{\Omega^{c}} + Z_{\Omega^{c}}, \eeq and
$K_{\Omega}$ is the conditional covariance matrix of $X_{\Omega}$
given $X_{\Omega^{c}}$.

It is also well-known (see Lemma $6.6$ in \cite{ADT09}) that restricting to the product distribution $Q_{p}=\mc{CN}(0,I)$, only leads to
relaxation of the mutual information term in \eqref{eq:cut-setGauss1} by at most
$\text{min}(|\Omega|,|\Omega^{c}|)$, and therefore \beqa
\bar{C}_{\mc{J}}^{\text{Gauss}} (Q) &\leq& \bar{C}_{\mc{J}}^{\text{Gauss}} (Q_{p}) + |\mc{V}|/2 \nn\\
&=& \min_{\Omega \in \Lambda_{\mc{J}}} \log \vert  I +
H_{\Omega,\Omega^{c}} H_{\Omega,\Omega^{c}}^{*} \vert + |\mc{V}|/2.
\eeqa

The following proposition shows that every cut in the Gaussian network is within a a constant gap to the cut in the DSN.
\bprop
For every $\Omega\in\Lambda_{\mc{J}}$, there exists a product distribution $Q_{p}^{\text{DSN}} = \prod_{v\in\mc{V}} p(X_{v}^{\text{DSN}})$ such that,
\beq
\log \vert  I + H_{\Omega,\Omega^{c}} H_{\Omega,\Omega^{c}}^{*} \vert \leq I \lp Y_{\Omega^{c}}^{\text{DSN}}; X_{{\Omega}}^{\text{DSN}} | X_{\Omega^{c}}^{\text{DSN}} \rp + O(|\mc{V}|), \label{eq:prod_form}
\eeq
\eprop
\bpf
We will reduce the jointly Gaussian vector $X$ to a vector that is valid in the DSN model and show that the reduction only leads to a $O(|\mc{V}|)$ loss in mutual information.
\begin{enumerate}
\item Note that
\beq \log \vert  I + H_{\Omega,\Omega^{c}} H_{\Omega,\Omega^{c}}^{*} \vert = I \lp H_{\Omega,\Omega^{c}} X_{\Omega} + Z_{\Omega^{c}}; X_{{\Omega}} \rp, \eeq
where $X_{v} \sim \text{i.i.d.}~\mc{CN}(0,1)$. Throughout the rest of this proof, we will drop the subscripts $\Omega$ and $\Omega^{c}$ for convenience.
It is shown in the proof of Theorem 4.1 in \cite{AK09} that if we restrict $X_{v}$ to only the fractional part, the loss in mutual information can be bounded.
We present a quick sketch here for the sake of completeness.
Let $\bar{X}_{v} \df X_{v} - \Lbr X_{v} \Rbr$ denote the fractional part. Then
\beqa
I \lp H X + Z; X \rp &\leq& I \lp H\bar{X}+Z, H\Lbr X\Rbr \rp \\
& \leq & I(H\bar{X}+Z; \bar{X}) + H(H\Lbr X\Rbr) \\
& \leq & I(H\bar{X}+Z; \bar{X}) + \sum_{v\in\Omega}H(\Lbr X\Rbr) \\
& \leq & I(H\bar{X}+Z; \bar{X}) + 4 |\Omega|.
\eeqa
Note that $\bar{X}_{v} \in [-1/2,1/2)$.

\item Next, from \cite{ADT09}(Lemma 7.2), it follows that
\beqa I(H\bar{X}+Z; \bar{X}) &\leq& I(\Lbr H\bar{X} \Rbr; \bar{X}) + 19 |\Omega^{c}|.
\eeqa
Here $\Lbr H\bar{X} \Rbr$ corresponds to the output of the DSN when $\bar{X}$ is the input, however $\bar{X}$ still takes values in a  continuous space and is not a
permissible input in the DSN.

\item Since the reduced channel given by $Y = \Lbr H \bar{X} \Rbr$ is deterministic and further the output $Y$ takes values in a finite set, we can
also restrict $\bar{X}_{v}$ to take values from a finite set within $[-1/2,1/2)$ without any loss of mutual information. Let us call this vector $X^{\text{DSN}}$,
\beqa I(\Lbr H\bar{X} \Rbr; \bar{X}) = I(\Lbr H{X}^{\text{DSN}} \Rbr; {X}^{\text{DSN}}). \eeqa

\end{enumerate}

We have therefore showed that
\beq \log \vert  I + H_{\Omega,\Omega^{c}} H_{\Omega,\Omega^{c}}^{*} \vert \leq I \lp Y_{\Omega^{c}}^{\text{DSN}}; X_{{\Omega}}^{\text{DSN}} | X_{\Omega^{c}}^{\text{DSN}} \rp + 23 |\mc{V}|.\eeq This completes the proof of proposition.
\epf

\end{document}

%% file: newcommand.tex
\newcommand{\nn}{\nonumber}
\newcommand{\bc}{\begin{center}}
\newcommand{\ec}{\end{center}}
\newcommand{\bfl}{\begin{flushleft}}
\newcommand{\efl}{\end{flushleft}}

\newcommand{\beqa}{\begin{eqnarray}}
\newcommand{\eeqa}{\end{eqnarray}}
\newcommand{\beqan}{\begin{eqnarray*}}
\newcommand{\eeqan}{\end{eqnarray*}}
\newcommand{\beq}{\begin{equation}}
\newcommand{\eeq}{\end{equation}}
\newcommand{\beit}{\begin{itemize}}
\newcommand{\eeit}{\end{itemize}}

\newcommand{\lbr}{\left \{ }
\newcommand{\rbr}{\right \} }
\newcommand{\Lbr}{\left [}
\newcommand{\Rbr}{\right ]}
\newcommand{\lp}{\left (}
\newcommand{\rp}{\right )}
\newcommand{\df}{\stackrel{{\rm def}}{=}}

\newcommand{\eqo}{\stackrel{\mbf{.}}{=}}
\newcommand{\lto}{\stackrel{\mbf{.}}{<}}
\newcommand{\gto}{\stackrel{\mbf{.}}{>}}

\newcommand{\prob}{{\mathbb{P}}}

\newcommand{\mc}{\mathcal}
\newcommand{\mb}{\mathbb}
\newcommand{\mf}{\mathfrak}
\newcommand{\mbf}{\mathbf}

\newcommand{\bit}{\begin{itemize}}
\newcommand{\eit}{\end{itemize}}
\newcommand{\ben}{\begin{enumerate}}
\newcommand{\een}{\end{enumerate}}

\newcommand{\blem}{\begin{lemma}}
\newcommand{\elem}{\end{lemma}}
\newcommand{\bthm}{\begin{theorem}}
\newcommand{\ethm}{\end{theorem}}
\newcommand{\bdefn}{\begin{definition}}
\newcommand{\edefn}{\end{definition}}
\newcommand{\bpf}{\begin{proof}}
\newcommand{\epf}{\end{proof}}
\newcommand{\bcor}{\begin{corollary}}
\newcommand{\ecor}{\end{corollary}}
\newcommand{\bprop}{\begin{proposition}}
\newcommand{\eprop}{\end{proposition}}

\newtheorem{proposition}{Proposition}
\newtheorem{lemma}{Lemma}
\newtheorem{theorem}{Theorem}

\newtheorem{remark}{Remark}
\newtheorem{corollary}{Corollary}

%% file: figs/network.pstex_t
\begin{picture}(0,0)%
\includegraphics{network.pstex}%
\end{picture}%
\setlength{\unitlength}{3947sp}%
\begingroup\makeatletter\ifx\SetFigFont\undefined%
\gdef\SetFigFont#1#2#3#4#5{%
  \reset@font\fontsize{#1}{#2pt}%
  \fontfamily{#3}\fontseries{#4}\fontshape{#5}%
  \selectfont}%
\fi\endgroup%
\begin{picture}(7393,5218)(879,-5557)
\put(4501,-2011){\makebox(0,0)[lb]{\smash{{\SetFigFont{14}{16.8}{\rmdefault}{\mddefault}{\updefault}{\color[rgb]{0,0,0}$R_1$}%
}}}}
\put(3301,-3361){\makebox(0,0)[lb]{\smash{{\SetFigFont{14}{16.8}{\rmdefault}{\mddefault}{\updefault}{\color[rgb]{0,0,0}$S$}%
}}}}
\put(2851,-5011){\makebox(0,0)[lb]{\smash{{\SetFigFont{14}{16.8}{\rmdefault}{\mddefault}{\updefault}{\color[rgb]{0,0,0}$R_3$}%
}}}}
\put(901,-1111){\makebox(0,0)[lb]{\smash{{\SetFigFont{14}{16.8}{\rmdefault}{\mddefault}{\updefault}{\color[rgb]{0,0,0}$D_2$}%
}}}}
\put(7351,-2161){\makebox(0,0)[lb]{\smash{{\SetFigFont{14}{16.8}{\rmdefault}{\mddefault}{\updefault}{\color[rgb]{0,0,0}$D_1$}%
}}}}
\put(4651,-5461){\makebox(0,0)[lb]{\smash{{\SetFigFont{14}{16.8}{\rmdefault}{\mddefault}{\updefault}{\color[rgb]{0,0,0}$D_3$}%
}}}}
\put(5551,-4111){\makebox(0,0)[lb]{\smash{{\SetFigFont{14}{16.8}{\rmdefault}{\mddefault}{\updefault}{\color[rgb]{0,0,0}$R_4$}%
}}}}
\put(1651,-2311){\makebox(0,0)[lb]{\smash{{\SetFigFont{14}{16.8}{\rmdefault}{\mddefault}{\updefault}{\color[rgb]{0,0,0}$R_2$}%
}}}}
\end{picture}%

%% file: figs/BC-Relay.pstex_t
\begin{picture}(0,0)%
\includegraphics{BC-Relay.pstex}%
\end{picture}%
\setlength{\unitlength}{3947sp}%
\begingroup\makeatletter\ifx\SetFigFont\undefined%
\gdef\SetFigFont#1#2#3#4#5{%
  \reset@font\fontsize{#1}{#2pt}%
  \fontfamily{#3}\fontseries{#4}\fontshape{#5}%
  \selectfont}%
\fi\endgroup%
\begin{picture}(4104,2904)(2149,-4213)
\put(5851,-4036){\makebox(0,0)[lb]{\smash{{\SetFigFont{14}{16.8}{\rmdefault}{\mddefault}{\updefault}{\color[rgb]{0,0,0}$D_2$}%
}}}}
\put(2326,-2836){\makebox(0,0)[lb]{\smash{{\SetFigFont{14}{16.8}{\rmdefault}{\mddefault}{\updefault}{\color[rgb]{0,0,0}S}%
}}}}
\put(3376,-1636){\makebox(0,0)[lb]{\smash{{\SetFigFont{14}{16.8}{\rmdefault}{\mddefault}{\updefault}{\color[rgb]{0,0,0}$R_1$}%
}}}}
\put(3451,-4036){\makebox(0,0)[lb]{\smash{{\SetFigFont{14}{16.8}{\rmdefault}{\mddefault}{\updefault}{\color[rgb]{0,0,0}$R_2$}%
}}}}
\put(5851,-1636){\makebox(0,0)[lb]{\smash{{\SetFigFont{14}{16.8}{\rmdefault}{\mddefault}{\updefault}{\color[rgb]{0,0,0}$D_1$}%
}}}}
\end{picture}%

%% file: figs/codingBlocks.pstex_t
\begin{picture}(0,0)%
\includegraphics{codingBlocks.pstex}%
\end{picture}%
\setlength{\unitlength}{3947sp}%
\begingroup\makeatletter\ifx\SetFigFont\undefined%
\gdef\SetFigFont#1#2#3#4#5{%
  \reset@font\fontsize{#1}{#2pt}%
  \fontfamily{#3}\fontseries{#4}\fontshape{#5}%
  \selectfont}%
\fi\endgroup%
\begin{picture}(10888,1714)(1157,-3269)
\put(10051,-3211){\makebox(0,0)[lb]{\smash{{\SetFigFont{12}{14.4}{\rmdefault}{\mddefault}{\updefault}{\color[rgb]{0,0,0}${f}_{\mc{V}_R}(T_3)$}%
}}}}
\put(10201,-1711){\makebox(0,0)[lb]{\smash{{\SetFigFont{12}{14.4}{\rmdefault}{\mddefault}{\updefault}{\color[rgb]{0,0,0}$t_3=T_3$}%
}}}}
\put(2251,-3211){\makebox(0,0)[lb]{\smash{{\SetFigFont{12}{14.4}{\rmdefault}{\mddefault}{\updefault}{\color[rgb]{0,0,0}${f}_{\mc{V}_R}(1)$}%
}}}}
\put(1276,-2386){\makebox(0,0)[lb]{\smash{{\SetFigFont{14}{16.8}{\rmdefault}{\mddefault}{\updefault}{\color[rgb]{0,0,0}$\vec{X}_v$}%
}}}}
\put(1201,-1936){\makebox(0,0)[lb]{\smash{{\SetFigFont{12}{14.4}{\rmdefault}{\mddefault}{\updefault}{\color[rgb]{0,0,0}$t_2=1$}%
}}}}
\put(3601,-1936){\makebox(0,0)[lb]{\smash{{\SetFigFont{12}{14.4}{\rmdefault}{\mddefault}{\updefault}{\color[rgb]{0,0,0}$t_2=T_2$}%
}}}}
\put(2251,-1711){\makebox(0,0)[lb]{\smash{{\SetFigFont{12}{14.4}{\rmdefault}{\mddefault}{\updefault}{\color[rgb]{0,0,0}$t_3=1$}%
}}}}
\end{picture}%

%% file: figs/outerCode-BC.pstex_t
\begin{picture}(0,0)%
\includegraphics{outerCode-BC.pstex}%
\end{picture}%
\setlength{\unitlength}{3947sp}%
\begingroup\makeatletter\ifx\SetFigFont\undefined%
\gdef\SetFigFont#1#2#3#4#5{%
  \reset@font\fontsize{#1}{#2pt}%
  \fontfamily{#3}\fontseries{#4}\fontshape{#5}%
  \selectfont}%
\fi\endgroup%
\begin{picture}(4704,2904)(1549,-4213)
\put(3301,-2761){\makebox(0,0)[lb]{\smash{{\SetFigFont{14}{16.8}{\rmdefault}{\mddefault}{\updefault}{\color[rgb]{0,0,0}$f_{\mc{V}_R}, g_j(.)$}%
}}}}
\put(5776,-4036){\makebox(0,0)[lb]{\smash{{\SetFigFont{14}{16.8}{\rmdefault}{\mddefault}{\updefault}{\color[rgb]{0,0,0}$\vec{Y}_{D_2}$}%
}}}}
\put(5776,-1636){\makebox(0,0)[lb]{\smash{{\SetFigFont{14}{16.8}{\rmdefault}{\mddefault}{\updefault}{\color[rgb]{0,0,0}$\vec{Y}_{D_1}$}%
}}}}
\put(1651,-2836){\makebox(0,0)[lb]{\smash{{\SetFigFont{14}{16.8}{\rmdefault}{\mddefault}{\updefault}{\color[rgb]{0,0,0}$\vec{X}_S$}%
}}}}
\end{picture}%

%% file: figs/reciprocity1.pstex_t
\begin{picture}(0,0)%
\includegraphics{reciprocity1.pstex}%
\end{picture}%
\setlength{\unitlength}{3947sp}%
\begingroup\makeatletter\ifx\SetFigFont\undefined%
\gdef\SetFigFont#1#2#3#4#5{%
  \reset@font\fontsize{#1}{#2pt}%
  \fontfamily{#3}\fontseries{#4}\fontshape{#5}%
  \selectfont}%
\fi\endgroup%
\begin{picture}(7372,3870)(586,-4651)
\put(3451,-1636){\makebox(0,0)[lb]{\smash{{\SetFigFont{14}{16.8}{\rmdefault}{\mddefault}{\updefault}{\color[rgb]{0,0,0}$R_1$}%
}}}}
\put(2326,-2836){\makebox(0,0)[lb]{\smash{{\SetFigFont{14}{16.8}{\rmdefault}{\mddefault}{\updefault}{\color[rgb]{0,0,0}D}%
}}}}
\put(3451,-4036){\makebox(0,0)[lb]{\smash{{\SetFigFont{14}{16.8}{\rmdefault}{\mddefault}{\updefault}{\color[rgb]{0,0,0}$R_2$}%
}}}}
\put(5851,-1636){\makebox(0,0)[lb]{\smash{{\SetFigFont{14}{16.8}{\rmdefault}{\mddefault}{\updefault}{\color[rgb]{0,0,0}$S_1$}%
}}}}
\put(5851,-4036){\makebox(0,0)[lb]{\smash{{\SetFigFont{14}{16.8}{\rmdefault}{\mddefault}{\updefault}{\color[rgb]{0,0,0}$S_2$}%
}}}}
\end{picture}%

%% file: figs/reciprocity2.pstex_t
\begin{picture}(0,0)%
\includegraphics{reciprocity2.pstex}%
\end{picture}%
\setlength{\unitlength}{3947sp}%
\begingroup\makeatletter\ifx\SetFigFont\undefined%
\gdef\SetFigFont#1#2#3#4#5{%
  \reset@font\fontsize{#1}{#2pt}%
  \fontfamily{#3}\fontseries{#4}\fontshape{#5}%
  \selectfont}%
\fi\endgroup%
\begin{picture}(7297,3870)(586,-4651)
\put(2326,-2836){\makebox(0,0)[lb]{\smash{{\SetFigFont{14}{16.8}{\rmdefault}{\mddefault}{\updefault}{\color[rgb]{0,0,0}S}%
}}}}
\put(3451,-4036){\makebox(0,0)[lb]{\smash{{\SetFigFont{14}{16.8}{\rmdefault}{\mddefault}{\updefault}{\color[rgb]{0,0,0}$R_2$}%
}}}}
\put(5851,-1636){\makebox(0,0)[lb]{\smash{{\SetFigFont{14}{16.8}{\rmdefault}{\mddefault}{\updefault}{\color[rgb]{0,0,0}$D_1$}%
}}}}
\put(5851,-4036){\makebox(0,0)[lb]{\smash{{\SetFigFont{14}{16.8}{\rmdefault}{\mddefault}{\updefault}{\color[rgb]{0,0,0}$D_2$}%
}}}}
\put(3451,-1636){\makebox(0,0)[lb]{\smash{{\SetFigFont{14}{16.8}{\rmdefault}{\mddefault}{\updefault}{\color[rgb]{0,0,0}$R_1$}%
}}}}
\end{picture}%

%% file: figs/coordination.pstex_t
\begin{picture}(0,0)%
\includegraphics{coordination.pstex}%
\end{picture}%
\setlength{\unitlength}{3947sp}%
\begingroup\makeatletter\ifx\SetFigFont\undefined%
\gdef\SetFigFont#1#2#3#4#5{%
  \reset@font\fontsize{#1}{#2pt}%
  \fontfamily{#3}\fontseries{#4}\fontshape{#5}%
  \selectfont}%
\fi\endgroup%
\begin{picture}(4104,2904)(2149,-4213)
\put(3451,-1636){\makebox(0,0)[lb]{\smash{{\SetFigFont{14}{16.8}{\rmdefault}{\mddefault}{\updefault}{\color[rgb]{0,0,0}$R_1$}%
}}}}
\put(2326,-2836){\makebox(0,0)[lb]{\smash{{\SetFigFont{14}{16.8}{\rmdefault}{\mddefault}{\updefault}{\color[rgb]{0,0,0}S}%
}}}}
\put(3451,-4036){\makebox(0,0)[lb]{\smash{{\SetFigFont{14}{16.8}{\rmdefault}{\mddefault}{\updefault}{\color[rgb]{0,0,0}$R_2$}%
}}}}
\put(5851,-1636){\makebox(0,0)[lb]{\smash{{\SetFigFont{14}{16.8}{\rmdefault}{\mddefault}{\updefault}{\color[rgb]{0,0,0}$D_1$}%
}}}}
\put(5851,-4036){\makebox(0,0)[lb]{\smash{{\SetFigFont{14}{16.8}{\rmdefault}{\mddefault}{\updefault}{\color[rgb]{0,0,0}$D_2$}%
}}}}
\put(2626,-2011){\makebox(0,0)[lb]{\smash{{\SetFigFont{14}{16.8}{\rmdefault}{\mddefault}{\updefault}{\color[rgb]{0,0,0}$\infty$}%
}}}}
\put(2626,-3511){\makebox(0,0)[lb]{\smash{{\SetFigFont{14}{16.8}{\rmdefault}{\mddefault}{\updefault}{\color[rgb]{0,0,0}$\infty$}%
}}}}
\end{picture}%